\documentclass[aps,prd,nopacs,floatfix,notitlepage,superscriptaddress,nofootinbib,twocolumn,a4paper,longbibliography]{revtex4-1}
\usepackage{amsfonts,amsmath,units,wasysym,epsfig,graphicx,verbatim,color,subfigure,graphicx,bm,mathrsfs,lipsum,hyperref}
\usepackage{wrapfig,comment,booktabs}
\usepackage[utf8]{inputenc}
\usepackage[normalem]{ulem}  
\usepackage{stfloats}    
\usepackage{placeins}     
\usepackage[dvipsnames]{xcolor}

\newcommand{\rn}[1]{\textsf{\color{blue}{ #1}}}

\newcommand{\rkb}[1]{\textsf{\color{brown}{ #1}}}

\begin{document}

\newcommand{\SNU}{Department of Physics, School of Natural Sciences,
Shiv Nadar Institution of Eminence, Greater Noida 201314, Uttar Pradesh, India}

\newcommand{\USAL}{Departamento de F\'isica Fundamental and IUFFyM, Universidad de Salamanca, Plaza de la Merced S/N, E-37008 Salamanca, Spain}

\newcommand{\Uliege}{Space Sciences, Technologies and Astrophysics Research (STAR) Institute, Universit   \'e de Li\`ege, B\^at. B5a, 4000 Li\`ege, Belgium}

\newcommand{\IITD}{Department of Physics, Indian Institute of Technology Delhi, New Delhi 110016, India}
\title{Implication of multimessenger observations on the relativistic mean-field equation of state of dense nuclear matter and skin thickness of nuclei}

\author{Rahul Kumar}
\email{rahulkumarrahul645@gmail.com}
\email{phz258498@iitd.ac.in}
\affiliation{\IITD}\affiliation{\SNU}

\author{Prasanta Char}
\email{prasanta.char@usal.es}
\affiliation{\USAL}\affiliation{\Uliege}

\author{Rana Nandi}
\email{rananandi@iitd.ac.in}
\affiliation{\IITD}\affiliation{\SNU}

\begin{abstract}
The composition and properties of infinite nuclear matter under extreme conditions of temperature and pressure remain incompletely understood. In this work, we constrain the equation of state (EoS) of nuclear matter$-$constructed within the framework of the Relativistic Mean Field (RMF) model$-$by combining results from chiral effective field theory and multimessenger observations of neutron stars. Using the saturation properties of nuclear matter, we generate a wide ensemble of EoS, which are subsequently constrained within a Bayesian framework. The resulting posterior distributions provide tight bounds on both the saturation parameters and the coupling constants of the RMF model.
The radius of a $1.4\,M_\odot$ neutron star is tightly constrained to $12.471_{-0.394}^{+0.427}\,\mathrm{km}$, while the maximum mass reaches $2.156_{-0.120}^{+0.175}\,M_\odot$. Furthermore, our analysis reveals that the $\omega$–$\rho$ coupling, which governs the density dependence of the symmetry energy, becomes increasingly significant under successive astrophysical constraints.
Finally, the predicted $\rm r_\mathrm{skin}$ of $^{48}\mathrm{Ca}$ agrees well with the CREX measurement, whereas that of $^{208}\mathrm{Pb}$ remains in tension with PREX-II. In contrast to earlier studies, we do not observe a clear correlation between the $\rm r_\mathrm{skin}$ of $^{208}$Pb and the symmetry energy slope parameter $L$. 
\end{abstract}

\maketitle

\section{Introduction}

Neutron stars (NS) are among the most extreme and fascinating objects in the universe, providing a natural laboratory for studying matter under conditions of immense density and pressure.  The mass of a canonical NS is around $1.4 \, M_\odot$, contained within a radius of about 10 km. This results in matter being compressed to densities several times higher than the normal nuclear density ($\rho_0$) \cite{Glendenning:1997wn}.
The composition and properties of matter at such extreme densities are governed by the equation of state (EoS), which remains uncertain because of the lack of precise theoretical and observational constraints. In principle, the behavior of nuclear matter can be derived directly from quantum chromodynamics (QCD), the fundamental theory of strong interactions. However, QCD becomes highly non-perturbative at low energies and finite baryon densities, making direct, first-principles calculations of nuclear systems too difficult and complex with current techniques. 

To overcome this challenge, a wide range of theoretical approaches have been developed to model the EoS of nuclear matter across densities relevant to finite nuclei, neutron stars, and core-collapse supernovae \cite{Oertel:2016bki}. 
Ab initio methods aim to solve the nuclear many-body problem starting from realistic two- and three-nucleon forces derived from chiral effective field theory ($\chi$-EFT) \cite{Tews:2012fj,Hebeler:2013nza,Lynn:2015jua, Drischler:2017wtt, Huth:2020ozf}. These methods have been used to compute the EoS of neutron-rich matter at densities up to $\sim 2\rho_0$.
However, they are less reliable at higher densities because of the breakdown of the EFT expansion and the increasing importance of unresolved short-range physics.

To go beyond the limitations of ab initio methods, various phenomenological models have been developed, designed to reproduce experimental observables in finite nuclei while also describing the matter at extreme densities as those found in NS cores. These include both non-relativistic energy density functionals, such as Skyrme and Gogny interactions, and relativistic density functionals based on the meson exchange Lagrangian \cite{Dutra:2012mb,Dutra:2014qga,Sun:2023xkg}. Phenomenological models also require significantly less computational resources compared to ab initio methods.

Among the relativistic frameworks, the Relativistic Mean-Field (RMF) model, originating from the Walecka model and extended to include nonlinear and cross-coupling terms, has been especially successful\cite{Glendenning:1997wn,WALECKA1974491, Serot1992}. RMF models are Lorentz covariant and naturally extendable to finite temperatures, suitable for supernova and NS merger simulations.  

In some modern variants, density-dependent couplings are introduced as an alternative to adding higher-order meson terms \cite{Fuchs:1995as, Typel:1999yq, Lalazissis:2005de, Typel:2018cap, Gogelein:2007qa}. Then there are meta-modeling approaches that employ a model-agnostic expansion of the EoS around nuclear saturation density \cite{Margueron:2017eqc, Margueron:2017lup}. Both non-relativistic and relativistic variants of meta-modeling exist and are often used within Bayesian frameworks to study correlations between nuclear matter parameters and neutron star observables \cite{universe7100373, Mondal:2022cva, Traversi:2020aaa, Zhu:2022ibs,Malik:2022zol, Beznogov:2022rri, Beznogov:2024vcv, Char:2023fue, Scurto:2024ekq, Passarella:2025zqb, Biswas:2019ifs, Biswas:2020xna}.

Moreover, there are several approaches that utilize flexible, semi-empirical parameterizations of the EoS, designed to explore a wide range of behaviors while satisfying physical constraints. These include piecewise polytropes \cite{Read:2008iy}, spectral representations \cite{Lindblom:2010bb, Lindblom:2012zi, Lindblom:2013kra} and speed of sound parameterizations \cite{Greif:2018njt}. Machine learning techniques have also gained significant attention in recent years for EoS inference, offering flexible, data-driven models that can capture complex behaviors without relying on fixed functional forms \cite{Landry:2018prl,Essick:2019ldf,Soma:2022qnv, Han:2022sxt, Zhou:2023cfs, Cuceu:2024hpq}. Although still in development, these methods show promise in capturing complex EoS behaviors that may not be easily represented in traditional parametric forms.

A variety of observational and theoretical inputs are essential for constraining the nuclear equation of state. At low densities, the chiral effective field theory ($\chi$-EFT) provides reliable predictions, anchoring the EoS near saturation density. Recent advancements in astrophysical observations have significantly improved our ability to probe the internal structure of NS and, consequently, the EoS governing dense matter. At higher densities, gravitational wave observations from binary neutron star mergers (e.g., GW170817) constrain the tidal deformability \cite{LIGOScientific:2017ync, LIGOScientific:2018hze, LIGOScientific:2018cki}, while NICER measurements of pulsars such as PSR J0030+0451\cite{Riley:2019yda,Miller:2019cac} and PSR J0740+6620\cite{Miller:2021qha,Riley:2021pdl} yield simultaneous mass–radius estimates. Enhanced observational techniques and advanced data analysis methods have accelerated NICER's mass-radius measurements of pulsars. In the last year alone, NICER has delivered three new radius measurements (PSR J0437-4715\cite{Choudhury_2024}, PSR J1231-1411\cite{Salmi:2024bss}, and PSR J0614-3329\cite{Mauviard:2025dmd}).

Because mass-radius and tidal deformability map directly onto the EoS, these measurements tightly constrain its behavior over the relevant density range. Additionally, kilonova emissions from neutron star mergers provide emerging constraints by linking the EoS to the properties of ejected matter\cite{Lund:2024fjk}, and radio pulsar timing has confirmed neutron stars with masses above 2 $M_\odot$, setting firm lower limits on the stiffness of the EoS \cite{Antoniadis:2013pzd, Fonseca:2016tux, Fonseca:2021wxt}.

Beyond astrophysical data, theory also provides an anchor at asymptotically high densities, where quark degrees of freedom become relevant.  In this regime, perturbative QCD (pQCD) calculations can be used to estimate the EoS of cold, dense matter\cite{Gorda_2023}. Although pQCD is not applicable at neutron star densities, it provides an essential boundary condition for models attempting to extrapolate the EoS toward high-density limits.

An important laboratory observable that connects the nuclear structure to the dense matter EoS is the neutron skin thickness, defined as the difference between the neutron and proton root mean squared radii in a nucleus. It provides important insight into the iso-spin dependence of the nuclear force. This quantity is found to be sensitive to the density dependence of the symmetry energy, especially its slope parameter $L$ at nuclear saturation density \cite{Vinas:2013hua,Reinhard:2016mdi}. A larger $L$ corresponds to a stiffer symmetry energy, which tends to push the neutrons further out, leading to a thicker neutron skin.

Recent measurements from the PREX-II (hereafter PREX) and CREX experiments have provided model-independent determinations of the neutron skin in \(^{208}\mathrm{Pb}\) and \(^{48}\mathrm{Ca}\), respectively, using parity-violating electron scattering\cite{PREX:2021umo,CREX:2022kgg}. While PREX suggested a relatively thick neutron skin in \(^{208}\mathrm{Pb}\) pointing towards a larger value of $L$\cite{PREX:2021umo}, the CREX result for \(^{48}\mathrm{Ca}\) reported a rather smaller skin thickness, favoring a softer symmetry energy around saturation density\cite{CREX:2022kgg}. The tension between these two measurements has generated significant theoretical interest\cite{PhysRevC.109.035803, PhysRevLett.129.232501}, as it may point to a more complex density dependence of the symmetry energy or to limitations in existing nuclear models \cite{PhysRevC.110.055802}. Together, PREX and CREX provide complementary constraints on the EoS probing the isovector sector of nuclear interactions from both heavy and medium-mass nuclei. Since the symmetry energy and its slope influence the neutron star radii and the pressure of neutron-rich matter, accurate neutron skin measurements from  PREX and CREX provide constraints on the EoS that are complementary to astrophysical observations.

Given the growing number of theoretical, observational, and laboratory constraints, Bayesian inference has emerged as a powerful and widely adopted tool to extract meaningful constraints on the EoS. In this work, we use Bayesian parameter estimation to explore the parameter space and constrain the EoS in light of the latest astrophysical observations.
In particular, we use the uncertainties of the nuclear empirical parameters to generate a large sample of EoSs using a nucleonic RMF model. We impose further constraints coming from $\chi$-EFT calculations, and combined astrophysical observations from GW, radio, and X-ray observations. Finally, we use our posterior samples for the RMF coupling parameters to calculate the neutron skin thickness for $^{208}$Pb and $^{48}$Ca. Then, we compare the skin thickness distributions coming from our calculations with the experimental results from PREX and CREX. This is the first time the Bayesian posteriors for the RMF parameters, informed by ab initio calculations and the multimessenger observations, have been used to calculate the distribution of the neutron skin thickness.

This paper is organized as follows.  In the next section, we discuss the EoS model and its parameters. Section \ref{Constraints and Bayesian} describes the details of the Bayesian framework that we have adopted and various theoretical, experimental, and observational data that we have utilized to constrain the EoS parameters. In \autoref{Results and Discussion} we discuss our findings, and finally we conclude with \autoref{Conclusion}. In this article, we use natural units, i.e., $\hbar=c=1$.

\section{Formalism}
\label{Formalism}
\subsection{Equation of State}
\label{EoS-Section}
    We employ a nonlinear finite-range RMF model, which comes under the class of Walecka-type models. In these models, the baryons interact among themselves via the exchange of mesons, which in our case are the scalar-isoscalar ($\sigma$), the vector-isoscalar ($\omega$), and the vector-isovector ($\rho$) mesons. The Lagrangian density of the model is given by \cite{Glendenning:1997wn, Nandi:2020luz, Dutra:2014qga}:
\begin{widetext}
\begin{eqnarray}
 {\cal L}=&&\sum_N\bar{\psi}_N\left[\gamma^\mu\left(i\partial_\mu   - g_{\omega}  \omega_\mu - \frac{1}{2}g_\rho\bm{\tau\cdot\rho_\mu}\right) 
  -\left(M_N- g_{\sigma }\sigma \right)\right]\psi_N 
   + \frac{1}{2}\left(\partial_\mu\sigma\partial^\mu\sigma - m_\sigma^2\sigma^2\right)
  - \frac{1}{3}bM_N(g_\sigma\sigma)^3 - \frac{1}{4}c(g_\sigma\sigma)^4 \nonumber \\
  && -\frac{1}{4}\omega_{\mu\nu}\omega^{\mu\nu}
  + \frac{1}{2}m_\omega^2\omega_\mu\omega^\mu
  -\frac{1}{4}\bm{\rho_{\mu\nu}\cdot\rho^{\mu\nu}}
   + \frac{1}{2}m_\rho^2\bm{\rho_\mu\cdot\rho^\mu}
   +\Lambda_{\omega \rho}g_\omega^2g_\rho^2\omega_\mu\omega^\mu\bm{\rho_\mu\cdot\rho^\mu}
   \label{eq:Ld}
\end{eqnarray}
\end{widetext}
where, $\psi_N$ is the isospin doublet of nucleons. Starting from the above Lagrangian density, we calculate the EoS of the nucleonic matter, relevant for the NS core. However, a NS core also contains leptons ($e^-$ and $\mu^-$) that maintain beta equilibrium with nucleons. Therefore, we need to add their contributions to the EoS.

For the crust, we use the standard SLy4 EoS \cite{Douchin:2001sv}.
To determine the crust-core transition point, we employ the thermodynamic spinodal method, which identifies it  as the density at which uniform npe$\mu$ matter becomes unstable against cluster formation, i.e., where the incompressibility under constant potential, $K_\mu$, changes sign to become negative. We use the following expression for $K_\mu$ for npe$\mu$ matter \cite{Kubis:2006kb, Lattimer:2006xb}:
\begin{equation}
\begin{split}
    K_{\mu} &= \rho^2\frac{d^2E_0}{d\rho^2} + 2\rho\frac{dE_0}{d\rho} \\
    &+ \delta^2\left[\rho^2\frac{d^2E_{\text{sym}}}{d\rho^2} + 2\rho\frac{dE_{\text{sym}}}{d\rho} - 2E_{\text{sym}}^{-1}\left(\rho\frac{dE_{\text{sym}}}{d\rho}\right)^2 \right]\, ,
\end{split}
\end{equation}
where the symbols are defined in the next section. This bulk criterion checks
whether infinite, homogeneous matter is stable against infinitesimal,
infinite-wavelength density fluctuations, and it slightly
overestimates the transition density. The actual crust-core transition may occur at a lower density when the dynamical spinodal is used instead, which additionally accounts for finite-size effects (surface tension and the Coulomb interaction)  and the wavelength dependence of the density fluctuations \cite{Pethick:1994ge, Ducoin:2010as, Ducoin:2011fy}. When matching to the crustal EoS, we also verify that there is no pressure discontinuity or causality violation at the crust-core boundary.

\begin{table}[h]
  \centering
  \caption{Prior Distributions for the Six EoS Parameters}
  \small
  \begin{tabular}{l @{\hspace{35pt}} c @{\hspace{35pt}} c}
    \hline\hline
    Parameter           & Unit         & Prior Distribution \\
    \hline
    $\rho_0$            & fm$^{-3}$    & $\mathcal{N}(0.152,\,0.004^2)$       \\
    $E_0$               & MeV          & $\mathcal{N}(-16.17,\,0.36^2)$       \\
    $M^*/M_N$           & —            & $\mathcal{U}[0.60,\,0.80]$           \\
    $K$                 & MeV          & $\mathcal{U}[170,\,320]$            \\
    $E_{\rm sym}$       & MeV          & $\mathcal{U}[25,\,40]$              \\
    $L$                 & MeV          & $\mathcal{U}[30,\,130]$             \\
    \hline\hline
  \end{tabular}
  \label{tab:prior_distributions}
\end{table}

\subsection{Parameters}
\label{subsec: Parameters}

The energy per nucleon $E(\rho ,\delta )$ at nucleon density $\rho$, isospin asymmetry $\delta\equiv (\rho_n-\rho_p)/\rho$ is given by,
\begin{equation}\label{eos0}
E(\rho ,\delta )=E_0(\rho)+E_{\rm{sym}}(\rho )\cdot \delta ^{2} +\mathcal{O}(\delta^4),
\end{equation}
Here, $E_0(\rho)$ is the energy per nucleon in symmetric nuclear matter (SNM), and $E_{\rm{sym}}(\rho)$ is the symmetry energy which can be expanded in Taylor series about $\rho_0$ as:
\begin{align}
  E_{0}(\rho) &= E_0(\rho_0) + \frac{K}{2} \chi^2 + \dots \\
  E_{\rm{sym}}(\rho) &= E_{\rm{sym}}(\rho_0) + L\, \raisebox{0.4ex}{$\chi$} \nonumber \\
  &\quad + \frac{K_{\rm{sym}}}{2} \chi^2 + \dots
\end{align}
where the dimensionless parameter $\chi$ is defined as
\begin{equation*}
    \chi = \left(\frac{\rho - \rho_0}{3\rho_0}\right)
\end{equation*}
In the expression for $E_0(\rho)$, the quantity $K$ is the incompressibility of SNM at saturation density which is given by 
\begin{equation}
    K=9\rho_0^2\left(\frac{\partial^2 E_0(\rho)}{\partial\rho^2}\right)_{\!0}\, ,
\end{equation} 
where $()_{0}$ indicates that the quantity in bracket is evaluated at saturation density. The other three parameters in the expression for $E_{\rm{sym}}(\rho)$ are the magnitude $E_{\rm{sym}}(\rho_0)$, slope $L$ and curvature $K_{\rm{sym}}$ of nuclear symmetry energy at saturation density, respectively. The expressions for $L$ and $K_{\rm{sym}}$ are given by

\begin{equation}
 L=3\rho_0\left(\frac{\partial E_{\rm{sym}}(\rho)}{\partial\rho}\right)_{\!0}   
 \label{eq:slope}
\end{equation}
and
\begin{equation}
 K_{\rm{sym}}=9\rho_0^2\left(\frac{\partial^2 E_{\rm{sym}}(\rho)}{\partial\rho^2}\right)_{\!0}   
\end{equation}

We consider six saturation properties of nuclear matter: the saturation density $\rho_0$, energy per nucleon $E_0(\rho_0)$, the effective nucleon mass $M^*(=M_N-g_{\sigma} \sigma$), the incompressibility $K$, the symmetry energy $E_{\rm{sym}}(\rho_0)$ and the slope of the symmetry energy $L$. The six coupling constants of the theory ($g_{\sigma},g_{\omega},g_{\rho},b,c,\Lambda_{\omega \rho}$) can be related algebraically to these six saturation properties. Given a set of $\rho_0$, $E_0(\rho_0)$, $M^*$, and $K$, we calculate the couplings related to the isoscalar sector ($g_{\sigma},g_{\omega},b,$ and $c$) by following the prescription given in Ref. \cite{Glendenning:1997wn}.  
Once the couplings in the isoscalar sector have been fixed, the isovector couplings, $g_\rho$ and $\Lambda_{\omega \rho}$, can be determined from $E_\text{sym} (\rho_0)$ and $L$ as shown in \textcite{PhysRevC.90.044305} (see also \cite{Hornick:2018kfi}). To derive analytical expressions for these couplings, one begins with the density-dependent expression for the symmetry energy, which, for the Lagrangian density given in Eq.~(\ref{eq:Ld}), can be written as:
\begin{eqnarray}
 E_{\rm{sym}}(\rho) &=& \frac{k_{\rm F}^{2}}{6E_{\rm F} } +
 \frac{g_{\rho}^{2}\,\rho}{8m_{\rho}^{\ast 2}} \\
 &=& E_{{\rm sym},0}(\rho) +  E_{{\rm sym},1}(\rho)
 \label{SymEnergy}
\end {eqnarray}
where
\begin{equation}
    E_{{\rm sym},0}(\rho) = \frac{k_{\rm F}^{2}}{6E_{\rm F} },
\end{equation}
and 
\begin{equation}
    E_{{\rm sym},1}(\rho) = \frac{g_{\rho}^{2}\,\rho}{8m_{\rho}^{\ast 2}}
\end{equation}
with
\begin{equation}
  \frac{m_{\rho}^{\ast 2}}{g_{\rho}^{2}} \equiv 
  \frac{m_{\rho}^{2}}{g_{\rho}^{2}}+2\Lambda_{\omega \rho}g_\omega^2 \omega_0^2\,, \label{eq:ms_rho}
\end{equation}
represent the isoscalar and isovector contributions to the symmetry energy, respectively. Here $k_F=(1.5\pi^2\rho)^{1/3}$ is the Fermi momentum of the nucleon and $E_F=\sqrt{k_F^2+M^{*2}}$ is the corresponding Fermi energy.

Now writing $J=E_{\rm sym}(\rho_0)$ and using the definition of $L$ (eq. (\ref{eq:slope})) we can split them into isoscalar and isovector parts as:
\begin{equation}
 J = J_0 + J_1 \;\; {\rm and} \;\; L=L_0 + L_1 \, ,
\end{equation}
where
\begin{align}
    J_0 &= E_{\rm{sym},0}(\rho_0) = \left(\frac{k_{\rm F}^{2}}{6E_{\rm F}}\right)_{\!0}\, , \\
    J_1 &= E_{\rm{sym},1}(\rho_0) = \left(\frac{g_{\rho}^{2}\rho}{8m_{\rho}^{\ast 2}}\right)_{\!0} \, , \\
    L_0 &= 3 \rho_0\!\left(\frac{d E_{\rm{sym},0}}{d\rho}\right)_{\!0} \, ,\\
    L_1 &= 3 \rho_0\!\left(\frac{d E_{\rm{sym},1}}{d\rho}\right)_{\!0}\label{eq:L1}\, ,
\end{align}
Since $J_0$ is already known from the saturation properties related to the isoscalar sector, we can readily calculate $J_1$ with a given value of $J$ as:
\begin{equation}
 J_{1}  = J-J_{0} .
 \label{J1}
\end{equation}
A similar procedure can be followed to calculate $L_1$ with a given $L$:
\begin{equation}
L_1 = L - L_0
 \label{L1a}
\end{equation}
where
\begin{align}
 L_{0} & = 3 \rho_0\!\left(\frac{d E_{\rm{sym},0}}{d\rho}\right)_{\!\!0} \notag \\
         & = J_{0} \Bigg(1+\frac{M^{\ast 2}}{E_{\rm F}^2} 
                   \left[1-\!\frac{3\rho}{\,M^{\ast}}\!\left(\frac{\partial M^{\ast}}
                   {\partial\rho}\right)\right]\Bigg)_{\!0} \,,	                  
 \label{L0}
\end {align}
with
\begin{eqnarray*}
 \left(\frac{{\partial M^{\ast}}}{\partial\rho}\right)_{\!\!0} &=& -
 \Bigg[
 \frac{M^{\ast}}{E_{\rm F}} 
 \left(\frac{m_\sigma^{\ast 2}}{g_\sigma^{2}}+\rho_{\rm s}^{\,\prime}(M^{\ast})\right)^{-1}\Bigg]_{\!0}\, ,\\
  \frac{m_\sigma^{\ast 2}}{g_\sigma^{2}}&\equiv &\frac{m_\sigma^{2}}{g_\sigma^{2}}+ 2bM_Ng_\sigma\sigma + 3cg_\sigma^2\sigma^2\, , 
 \\
    \rho_{\rm s}^{\,\prime}(M^{\ast})  &=& 
 \left(\frac{\partial\rho_{\rm s}}{\partial M^{\ast}}\right) \notag \\
 &=& \frac{1}{\pi^{2}}
 \left[\frac{k_{\rm F}}{E_{\rm F}}(E_{\rm F}^2+2M^{\ast 2}) -
 3M^{\ast 2}\ln\left(\frac{k_{\rm F}+E_{\rm F}}{M^{\ast}}\right)\right] .
 \label{Rhos1}
\end{eqnarray*}
can again be determined from the isoscalar sector.
 
Once $L_1$ is known,  we can calculate $\Lambda_{\omega \rho}$ from eq.(\ref{eq:L1}) as:
\begin{align}
 L_{1} & = 3 \rho_0\!\left(\frac{d E_{\rm{sym},1}}{d\rho}\right)_{\!\!0} \notag \\
         &  = 3J_{1} 
                   \left[1-32\left(\frac{g_\omega^{2}}{m_\omega^{2}}\right)\!g_\omega \omega_0
                   \Lambda_{\omega \rho} J_{1}\right]_{0} \notag
\end{align}
Or,
\begin{equation}
    \Lambda_{\omega \rho} = \frac{3J_1-L_1}{96J_1^2(g_\omega/m_\omega)^2g_\omega\omega_0}
\end{equation}

Finally, we can determine $g_\rho$ by putting the value of $\Lambda_{\omega \rho}$ in eq. (\ref{eq:ms_rho}) as:
\begin{equation*}
 \frac{m_{\rho}^{2}}{g_{\rho}^{2}} =
 \frac{m_{\rho}^{\ast 2}}{g_{\rho}^{2}} - 2\Lambda_{\omega \rho}g_\omega^2 \omega_0^2 =
 \frac{\rho_0}{8J_{1}} - 2\Lambda_{\omega \rho}g_\omega^2 \omega_0^2 \,.
 \label{grho2}
\end {equation*}
Or,
\begin{equation}
    g_\rho^2 = \frac{m_\rho^2}{\rho_0/8J_1-2\Lambda_{\omega\rho}g_\omega^2\omega_0^2} \, .
\end{equation}

In this study, we adopted the following values for the masses of nucleons and mesons: $M_N=939$ MeV  $m_\sigma=550$ MeV, $m_\omega=782.5$ MeV, and $m_\rho=763$ MeV.


\section{Constraints and Bayesian Inference}
\label{Constraints and Bayesian}
\subsection{Bayesian Inference}
\label{Bayesian_Inference}
In the Bayesian framework, prior knowledge about model parameters is formally incorporated through a prior distribution, which is then updated using the likelihood of the observed data. The result is the posterior distribution, which is obtained via Bayes' theorem:
\begin{equation*}
    P(\theta | D) = \frac{P(D | \theta) P(\theta)}{\int P(D | \theta) P(\theta) d\theta} \, .
\end{equation*}
Here, $ P(\theta)$ is the prior probability, $P(D | \theta)$ is the likelihood function,  $P(D) = \int P(D | \theta) P(\theta) d\theta $ is the normalization constant also called the evidence, and $P(\theta | D)$ is the posterior probability of the model parameters given the dataset $D$. The evidence can be computed by integrating the numerator over the entire parameter space. But as the dimension of the parameter space increases, it becomes computationally inefficient. However, we can overcome this difficulty by adopting statistical computation techniques such as Markov Chain Monte Carlo (MCMC) or Nested Sampling. 

In this work, we're using Bayesian statistics for parameter estimation via Pymultinest\cite{Buchner:2014nha}, which is based on a nested sampling algorithm. While MCMC can struggle with complex or multi-modal distributions, nested sampling, by design, can more efficiently map out different regions of high likelihood, providing a better global view of the parameter space. This method not only provides the posterior distribution but also accurately computes the Bayesian evidence.

\subsection{Prior}
As discussed in \autoref{subsec: Parameters}, we have the six saturation properties of nuclear matter: the saturation density $\rho_0$, energy per baryon $E_0$, the effective nucleon mass $M^*$, the incompressibility $K$, the symmetry energy $E_{\text{sym}}(\rho_0)$ and the slope of the symmetry energy $L$ as parameters for our work. The ranges and the type of prior we're using is given in \autoref{tab:prior_distributions}. Since the first two parameters control mainly the low-density behavior of EoS, these two are well constrained from nuclear theory and experiments, and therefore, we use gaussian priors for them. For the remaining less-constrained parameters, we use uninformative priors between the specified ranges, i.e., we sample each parameter uniformly between its minimum and maximum value.

\subsection{Likelihood}
\label{likelihood}
We construct a joint likelihood from $\chi-$EFT, maximum mass, GW170817 and NICER constraints. The joint likelihood is a product of likelihood contributions from these four independent components and is given by,
\begin{equation}
    \mathcal{L}_{\rm tot} = \mathcal{L}_{\chi-\text{EFT}} \times \mathcal{L}_{M_{\rm max}} \times \mathcal{L}_{\rm NICER} \times \mathcal{L}_{\rm GW}\,.
\end{equation}

\subsubsection{Chiral Effective Field Theory ($\chi$-EFT)}

$\chi$-EFT is rooted in QCD and describes nucleon and pion interactions at low energies. For our work, we use constraints on the pressure of neutron-star matter derived from the $\chi$-EFT calculations of Drischler et al (\cite{Drischler:2015eba, Drischler:2017wtt, Drischler:2021kxf}), extended to beta-equilibrated matter via the formalism of Hebeler et al. \cite{Hebeler:2013nza} as implemented by Raaijmakers et al. \cite{Raaijmakers:2021uju,raaijmakers_2021_4696232}. These calculations provides a pressure band, defined by a lower bound $p_{\rm low}$ and an upper bound $p_{\rm up}$, for densities up to 1.12 times the nuclear saturation density.

At each density point we model the allowed pressure range as a Gaussian with mean $\bar{p} = (p_{\rm up} + p_{\rm low})/2$ and standard deviation $\sigma_p = (p_{\rm up} - p_{\rm low})/2$. The $\chi-$EFT likelihood is then the product of the pointwise Gaussian probabilities evaluated at the pressure of the EoS, $p_{\rm EoS}(\rho_i)$:
\begin{equation}
    \mathcal{L}_{\chi-\text{EFT}} = \prod_{i}
    \frac{1}{\sqrt{2\pi}\,\sigma_{p,i}}
    \exp\!\left[-\frac{\bigl(p_{\rm EoS}(\rho_i) - \bar{p}_i\bigr)^2}
                       {2\sigma_{p,i}^2}\right].
\end{equation}

\subsubsection{Maximum Mass}

The precise mass measurement of the neutron star PSR J0740+6620, with a mass of  $2.08 \pm 0.07 \, M_{\odot}$, currently represents the highest reliably determined mass of any known neutron star\cite{Riley:2021pdl}. This serves as a critical constraint on the EoS, as it sets a firm lower bound on the pressure required to support such a massive object against gravitational collapse. An EoS that becomes too soft at high densities would fail to support a star of this mass, and is therefore ruled out by this observation. As such, the existence of PSR J0740+6620 imposes one of the most stringent constraints on the stiffness of the EoS at supranuclear densities.

We require the EoS to support a maximum gravitational mass ($M_{\rm max}$) satisfying $M_{\rm max} \geq \mu_{{\rm J0740}} - 5\sigma_{{\rm J0740}}$ as a hard cut, where $\mu_{{\rm J0740}} = 2.08\,M_\odot$ and
$\sigma_{{\rm J0740}} = 0.07\,M_\odot$. EoS configurations that survive this cut are further weighted by the cumulative distribution function (CDF) of the PSR J0740+6620 mass measurement :
\begin{align}
    \mathcal{L}_{M_{\rm max}} 
    & = \int_{-\infty}^{M_{\rm max}} \frac{1}{\sqrt{2\pi}\,\sigma_{\rm J0740}} \exp\left[-\frac{(m - \mu_{\rm J0740})^2}{2\sigma_{\rm J0740}^2}\right] dm \notag \\
    & = \frac{1}{2}\left[ 1 + {\rm erf}\left(\frac{M_{\rm max} - \mu_{\rm J0740}}{\sqrt{2}\sigma_{\rm J0740}}\right)\right]\, ,
\end{align}

\subsubsection{Gravitational Wave Constraint (GW170817)}

On August 17, 2017, at 12:41:04 UTC, the Advanced LIGO and Advanced Virgo gravitational-wave detectors made their first observation of a binary neutron star inspiral \cite{LIGOScientific:2017vwq,LIGOScientific:2017ync, Abbott:2018exr, LIGOScientific:2018hze}. The masses and tidal deformabilities of the two component stars were estimated from this detection\footnote{\url{https://dcc.ligo.org/LIGO-P1800061/public}}. 

We use the low-spin posterior samples of the \textsc{PhenomPNRT} waveform model and construct a four-dimensional KDE, $\hat{f}{_{\rm GW}}(m_1, m_2, \Lambda_1, \Lambda_2)$, over the component masses and dimensionless tidal deformabilities. The primary component mass $m_1$ is drawn uniformly from  $[1.36, 1.61]\,M_\odot$, consistent with the GW170817 inference using the \textsc{PhenomPNRT} waveform model\cite{ LIGOScientific:2018hze}. The companion mass $m_2$ is then solved for analytically from the chirp-mass relation.

For each EoS generated in the Bayesian analysis, we simulate  $(N=100)$ binaries by sampling $m_{1}$ from the above range and computing the corresponding  $m_{2}$. The corresponding tidal deformabilities $\Lambda_{1}$ and $\Lambda_{2}$ are obtained by interpolating the $M-\Lambda$ sequence, which is obtained by solving the tidal perturbation equations\cite{Flanagan:2007ix,Hinderer:2007mb,Hinderer:2009ca} along with the Tolman-Oppenheimer-Volkoff (TOV) equations\cite{Tolman:1939jz,Oppenheimer:1939ne,Glendenning:1997wn} for the given EoS. 
The GW likelihood for the EoS is then given by:
\begin{equation}
    \mathcal{L}_{\rm GW} = \frac{1}{N}
    \sum_{i=1}^N
    \hat{f}{_{\rm GW}}\!\left(m_{1,i},\, m_{2,i},\,
    \Lambda(m_{1,i}),\, \Lambda(m_{2,i})\right)\,.
\end{equation}

\subsubsection{NICER Mass-Radius Constraints}

NICER provides mass and radius measurements for pulsars that provide strong constraints on the EoS. The headline 68\% credible intervals(CIs) for the five pulsar measurements are summarized in \autoref{tab:nicer_mr}. 
Although mass-radius data are available for five pulsars to date, we only incorporate the first three and the latest one in our likelihood calculations. We omit NICER 4 due to ambiguity in its mass-radius data. \cite{Salmi:2024bss}.

For each NICER source $s$, we construct a kernel density estimate (KDE) $\hat{f}_s(M,R)$ from the published posterior samples \cite{riley_2019_3386449, riley_2021_4697625} \cite{choudhury_2024_13766753,mauviard_2025_17380576}. We also uniformly sample masses $\{m_i^{(s)}\}$ within the range of NICER posterior. 
Now, for every sampled mass $\{m_i^{(s)}\}$, corresponding radius is obtained by interpolating the mass-radius $M-R$ sequence, which is obtained by solving the TOV\cite{Tolman:1939jz,Oppenheimer:1939ne,Glendenning:1997wn} equations with the help of the current EoS. The likelihood contribution from the source $s$ is then evaluated with the help of the KDE we constructed for that source and is given by: 
\begin{equation}
    \mathcal{L}_s = \frac{1}{N_s}\sum_{i=1}^{N_s}
    \hat{f}_s\!\left(m_i^{(s)},\, R\!\left(m_i^{(s)}\right)\right)\, ,
\end{equation}
where only those mass points $m_i^{(s)}$ that do not exceed the maximum mass of the current EoS are included. The combined NICER likelihood is given by: 
\begin{equation}
    \mathcal{L}_{\rm NICER} = \prod_{s}
    \mathcal{L}_s\,
\end{equation}
where $s \in \{{\rm J0030, J0740, J0437, J0614}\}$.

\subsubsection{Joint Log-Likelihood}
The joint likelihood is then passed to {\textsc{PyMultiNest}} as log-likelihood and is given by: 
\begin{equation}
    \ln\mathcal{L}_{\rm tot} =
    \ln\mathcal{L}_{\chi-\text{EFT}} +
    \ln\mathcal{L}_{M_{\rm max}} +
    \ln\mathcal{L}_{\rm NICER} +
    \ln\mathcal{L}_{\rm GW}\,.
\end{equation}
Any EoS that fails to converge numerically is assigned $\ln\mathcal{L}_{\rm tot} = -\infty$ and is excluded from the posterior. We set the number of live points equal to 8192, for our Bayesian runs. The acceptance rate ranges from 0.27 to 0.34,   the total number of likelihood evaluations ranges from around 146000 to around 248000. 
\begin{table}[h]
\centering
\caption{NICER mass and radius measurements}
\label{tab:nicer_mr}
\begin{tabular}{c c c c c}
\hline\hline
NICER & Pulsar & Mass($M_\odot$) & Radius(km) & Ref. \\ \hline 
1 & PSR J0030+0451 & $1.44^{+0.15}_{-0.14}$ & $13.02^{+1.24}_{-1.06}$ & \cite{Riley:2019yda, riley_2019_3386449} \\
2 & PSR J0740+6620 & $2.08 \pm 0.07$ & $13.7^{+2.6}_{-1.5}$ & \cite{Riley:2021pdl, riley_2021_4697625} \\
3 & PSR J0437--4715 & $1.418 \pm 0.037$ & $11.36^{+0.95}_{-0.63}$ & \cite{Choudhury_2024, choudhury_2024_13766753} \\
4${^\dagger}$ & PSR J1231--1411 & $1.04^{+0.05}_{-0.03}$ & $12.6 \pm 0.3$ & \cite{Salmi:2024bss, salmi_2024_13358349} \\
5 & PSR J0614--3329 & $1.44^{+0.06}_{-0.07}$ & $10.29^{+1.01}_{-0.86}$ & \cite{Mauviard:2025dmd, mauviard_2025_17380576} \\ [2pt]
\hline
\end{tabular}\vspace{4pt}
\raggedright\footnotesize{$^{*}$NICER numbers are shorthand labels used in this work for convenience; the Pulsar column lists the original name of the pulsars\\
${^\dagger}$ NICER 4 is not used in this work.}
\end{table}

\section{Results and Discussion}
\label{Results and Discussion}
We have used nuclear saturation properties as our parameters instead of the couplings in the Lagrangian. This choice offers several advantages. Saturation properties are physically meaningful and directly constrained by experimental data from finite nuclei, unlike the more abstract meson–nucleon and meson-meson  couplings. Moreover, this parametrization allows for a clearer connection between nuclear observables and the EoS, and is better suited for Bayesian inference, where physically motivated priors can be imposed and interpreted more directly.


We impose our observational and theoretical constraints in five sequential stages, each stage having its own set of constraints. At each stage a new constraint is added to the likelihood and an independent Bayesian analysis is performed, thereby obtaining the posterior set of parameters corresponding to that likelihood.

First, we generate posterior samples for all EoSs, that satisfy chiral effective field theory at low densities and reproduce at least the maximum observed neutron‑star mass (PSR J0740+6620). This defines our $\chi$-EFT+M$_{\rm max}$ baseline. Second, we further add to this likelihood the tidal deformability bounds from GW170817, and generate the new samples. Third, we jointly enforce the mass–radius posteriors from the first two NICER targets NICER 1 and 2. We do the same for the more recent measurement from NICER 3. Finally, we apply the latest NICER measurement NICER 5 to arrive at our most tightly constrained set. The same ordering as explained above also appears in the figure legends, allowing us to see precisely how each new piece of data refines our knowledge of the neutron‑star EoS.


We do not include NICER 4 (PSR J1231-1411) in our analysis due to the strong dependence of its inferred radius on the assumed priors. The results vary significantly based on whether informative or uninformative priors are used, with the radius ranging from $12.6\pm0.3$ km to $13.5^{+0.3}_{-0.5}$ km. Moreover, when broader radius priors are applied, the models fail to fit the data well. This sensitivity to prior assumptions, along with the model-dependent nature of the inferred geometry of the hot spots, makes the constraints from PSR J1231 less robust. It is worth noting, however, that some recent studies do incorporate PSR J1231 in their analysis \cite{LI2025139501, Li:2025oxi}, but we choose not to do it in this study.

\subsection{Saturation Properties}

\begin{figure*}[h]
    \centering
    \includegraphics[width=1\linewidth]{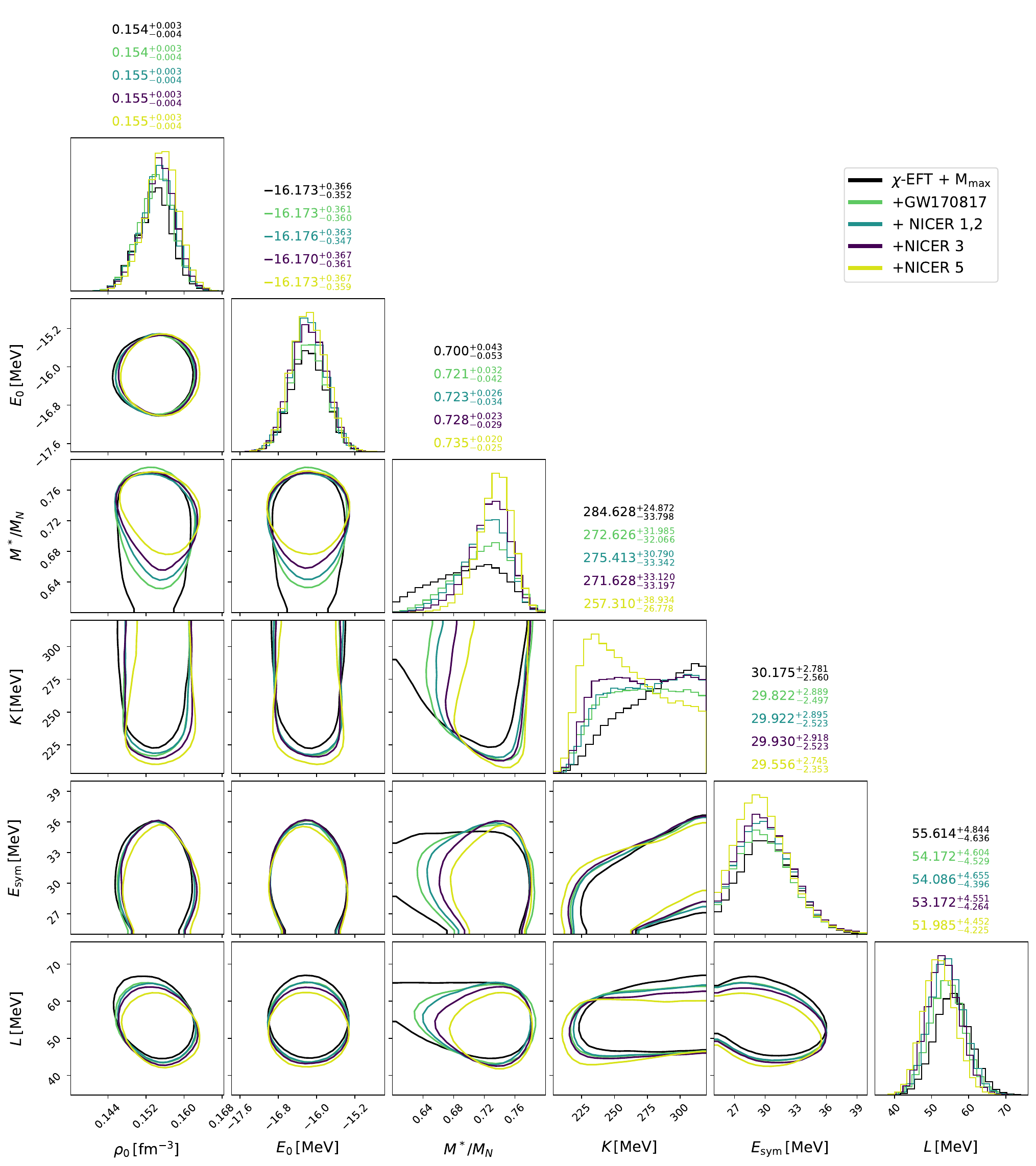}
    \caption{ Corner plot depicting the posterior distributions of the nuclear saturation parameters. The diagonal panels display the one-dimensional marginalized distributions for each parameter, while the off-diagonal panels show the 90\% contours for the two-dimensional joint probability density functions. The colored contours and distributions represent successive constraints applied in the order indicated by the legend. On top of each diagonal panels we also show the 68\% CI for each parameter in the order indicated by the legend. }
    \label{fig:corner_saturation}
\end{figure*}
We construct the joint probability density functions (PDFs) of the parameters {$\rho_0$, $E_0$, $M^*$, $K$, $E_{\text{sym}}$, $L$} using posterior samples obtained from the chosen prior distributions and the likelihood function. \autoref{fig:corner_saturation} shows the joint and marginalized posterior distributions of these parameters, with successive application of constraints from nuclear theory and astrophysical observations. The diagonal panels show the marginalized one-dimensional probability distributions for each parameter, while the off-diagonal panels represent the joint PDFs between pairs of parameters, illustrating their correlations.

\begin{table*}[h]
    \centering
    \caption{Posterior 90\% CI for the six EoS parameters}
    \small
    \begin{tabular}{l c c c c c c}
        \hline\hline
        Parameters & \hspace{10pt} Unit & \hspace{10pt} $\chi$-EFT+ $\text{M}_\text{max}$  & \hspace{10pt} +GW170817 & \hspace{10pt} + NICER 1,2 & \hspace{20pt} + NICER 3 & \hspace{10pt} +NICER 5\\
        \hline
        $\rho_0$   & \hspace{10pt} $\mathrm{fm}^{-3}$ 
                   & \hspace{10pt} $0.154_{-0.006}^{+0.006}$  
                   & \hspace{10pt} $0.154_{-0.006}^{+0.006}$ 
                   & \hspace{10pt} $0.155_{-0.006}^{+0.005}$ 
                   & \hspace{10pt} $0.155_{-0.006}^{+0.005}$ 
                   & \hspace{10pt} $0.155_{-0.006}^{+0.005}$ \\
        $E_0$      & \hspace{10pt} $\mathrm{MeV}$ 
                   & \hspace{10pt} $-16.173_{-0.587}^{+0.592}$ 
                   & \hspace{10pt} $-16.173_{-0.588}^{+0.594}$ 
                   & \hspace{10pt} $-16.176_{-0.572}^{+0.593}$ 
                   & \hspace{10pt} $-16.170_{-0.587}^{+0.590}$
                   & \hspace{10pt} $-16.173_{-0.587}^{+0.592}$ \\
        $M^*/M$    & \hspace{10pt} $-$ 
                   & \hspace{10pt} $0.700_{-0.079}^{+0.063}$ 
                   & \hspace{10pt} $0.721_{-0.069}^{+0.048}$ 
                   & \hspace{10pt} $0.723_{-0.061}^{+0.040}$ 
                   & \hspace{10pt} $0.728_{-0.053}^{+0.036}$
                   & \hspace{10pt} $0.735_{-0.044}^{+0.032}$ \\
        $K$        & \hspace{10pt} $\mathrm{MeV}$ 
                   & \hspace{10pt} $284.628_{-50.382}^{+31.984}$ 
                   & \hspace{10pt} $272.626_{-45.579}^{+42.440}$ 
                   & \hspace{10pt} $275.413_{-46.369}^{+40.108}$ 
                   & \hspace{10pt} $271.628_{-45.203}^{+43.530}$ 
                   & \hspace{10pt} $257.310_{-36.071}^{+54.706}$ \\
        $E_{\mathrm{sym}}$ & \hspace{20pt} $\mathrm{MeV}$ 
                   & \hspace{10pt} $30.175_{-3.949}^{+4.628}$ 
                   & \hspace{10pt} $29.822_{-3.774}^{+5.009}$ 
                   & \hspace{10pt} $29.922_{-3.820}^{+4.885}$ 
                   & \hspace{10pt} $29.930_{-3.832}^{+5.032}$ 
                   & \hspace{10pt} $29.556_{-3.583}^{+4.891}$ \\
        $L$        & \hspace{10pt} $\mathrm{MeV}$ 
                   & \hspace{10pt} $55.614_{-7.681}^{+8.062}$ 
                   & \hspace{10pt} $54.172_{-7.368}^{+7.763}$ 
                   & \hspace{10pt} $54.086_{-7.161}^{+7.778}$ 
                   & \hspace{10pt} $53.172_{-6.987}^{+7.536}$ 
                   & \hspace{10pt} $51.985_{-6.832}^{+7.463}$ \\
        \hline\hline
    \end{tabular}
    \label{tab:posterior_ranges_saturation_90CI}
\end{table*}
The 90\% posterior ranges of the nuclear saturation parameters are given in \autoref{tab:posterior_ranges_saturation_90CI}. We also show the 68\% CI for each parameter on top of each diagonal panel of \autoref{fig:corner_saturation}. From both sets of data, it is evident that the effective mass $M^*$, the symmetry energy $E_\text{sym}$, and its slope $L$ at saturation are remarkably constrained. Although the constraint on incompressibility $K$ at saturation is broader, it still shows a significant improvement relative to its chosen prior range. Overall, we can say that we have very tight constraints on all of our parameters.
The posterior median for $\rho_0$ sits at $\sim 0.155$ $\mathrm{fm}^{-3}$ across all datasets. The addition of NICER constraint shifts the median towards a little higher value but it still remains under the baseline uncertainty. Similarly the median value of $E_0$  remains stable at -16.17 MeV across all datasets changing narrowly within the baseline uncertainty. This result is consistent with existing literature across different models and constraint sets\cite{Passarella:2025zqb,Malik:2023mnx,Malik:2023mnx,Scurto:2024ekq,Traversi:2020aaa,Providencia:2023rxc,Arvikar:2025dwl}. The very stability of $\rho_0$ and $E_0$ demonstrates that astrophysical data to date-GW170817 and NICER-act as consistency checks on both of these saturation properties rather than refining them appreciably beyond what is known from nuclear theory and experiments.


The posterior median of the nucleon effective mass shifts to higher values when astrophysical data from GW170817 and NICER (1, 2, 3, and 5) are incorporated, increasing from 0.70 to around 0.74. 
Since a higher effective mass generally leads to a softer equation of state (EoS), this trend suggests a preference for softer EoSs under most constraint sets. Notably, NICER 5 yields the highest median effective mass, indicating the strongest preference for a soft EoS. In addition to the systematic shift in the median, the associated uncertainties are also reduced as astrophysical observations are applied.
This increase in effective mass was also found in a recent study \cite{Passarella:2025zqb}, where a density-dependent coupling for the $\rho$ meson was used instead of the constant coupling scheme we use here. Our result is also consistent with \textcite{Zhu:2022ibs} at 90\% CI and with \textcite{Arvikar:2025dwl} at 68\% CI, which use the same Lagrangian as ours. A similar Bayesian study \cite{Traversi:2020aaa} but without the $\omega-\rho$ coupling in the Lagrangian  got an even higher effective mass for their baseline case. The  inclusion of $\omega-\rho$ coupling gives an extra freedom to tune the slope of the symmetry energy.

For our baseline case, the incompressibility $K$ has a 90\% CI of $285^{+32}_{-50}$ MeV. This is largely consistent with the result of \textcite{Malik:2023mnx} for set 1, $257^{+36}_{-23}$ MeV, obtained with a hard cutoff for the maximum mass ($M_{\rm max}>2.0M_\odot$), an earlier $\chi$-EFT result\cite{Hebeler:2013nza}, and narrower prior ranges for saturation parameters including $K(230\pm40)$ MeV. We compare only with their set 1, as it has the lowest $\omega$-meson self-coupling, which we do not include. It softens the EoS at high densities \cite{Bodmer:1991hz, Mueller:1996pm, Horowitz:2001ya}.
Including the  GW170817 constraint shifts the median downward to 273 $\mathrm{MeV}$. Since a lower incompressibility generally leads to a softer EoS, this indicates that the tidal‐deformability measurements favors a softer EoS. When compared with the analysis of \textcite{Passarella:2025zqb}, where the maximum mass constraint was implemented differently, we still find considerable agreement between the 68\% CIs. Looking at the effect of each successive NICER dataset, combined NICER-1 and NICER-2 increase the median incompressibility, consistent with a stiffer EoS, whereas NICER-3 reduces it, pulling the overall median down to 272 MeV. The most recent NICER-5 release has the strongest impact, driving the median to its lowest value of 257 $\mathrm{MeV}$.  Our 90\% interval, obtained without imposing NICER-3 and NICER-5, coincides closely with that of \textcite{Scurto:2024ekq}, who used unified equations of state within an RMF model with density-dependent couplings and likewise did not have NICER-3 or NICER-5. However, their uncertainty is larger than ours.  We find similar agreement with other works \cite{Imam:2024gfh,Imam:2025lut,Zhu:2022ibs}.
Our results disagree with those of \textcite{Arvikar:2025dwl} at 68\% CI. This discrepancy likely arises from their incorporation of additional finite-nuclei and heavy-ion collision constraints, along with a factor that penalizes very stiff EoS. It is worth noting that the total uncertainty for the NICER-5 case remains comparable to that of baseline and previous multimessenger constraints. The inclusion of NICER-5  data also makes the posterior distribution more asymmetric, shifting it toward lower values.

In our analysis the symmetry energy $E_{\rm sym}$ evolves from a broad, flat prior of $25–40$ MeV to a tightly peaked posterior at around $30 \ \mathrm{MeV}$. In particular, our baseline constraint yields $E_{ \rm sym} \approx 30.2$ $\mathrm{MeV}$. Adding GW170817 and the NICER datasets shifts the median by no more than 0.6 MeV and leaves the credible bounds essentially intact.  
In other words, $\chi$-EFT and maximum mass alone impose very tight constraints on symmetry energy. Our results for $E_{\rm sym}$ agree well with those of previous studies \cite{Traversi:2020aaa, Passarella:2025zqb, Malik:2023mnx,Scurto:2024ekq, Zhu:2022ibs,Arvikar:2025dwl}.

The slope of the symmetry energy, $L$, is driven from a completely uninformative uniform prior over $30-130$ $\mathrm{MeV}$ to a sharply peaked posterior of $55.6^{+4.8}_{-4.6}$ MeV at $68\%$ CI as we impose our baseline constraint. Adding the GW170817 constraint shifts the median downwards, indicating it favors a slightly softer density dependence. NICER-1,2 does not have much effect on $L$, while NICER-3 reduces the median to $53.2$ $\mathrm{MeV}$. Finally NICER-5 shows a stronger effect decreasing the median to 52.0 $\mathrm{MeV}$. However, the baseline $68\%$ CI does not change upon application of the GW170817 and NICER constraints. Our results for $L$ agree well with \textcite{Passarella:2025zqb}, both with and without NICER-5, and with other recent works \cite{Zhu:2022ibs,Scurto:2024ekq, Malik:2023mnx,Arvikar:2025dwl}. Our results are also consistent with constraint set 3 and 4 of \textcite{Imam:2025lut}.


In the two‐dimensional panels of the corner plot, we show the 90\% credible contours to capture the parameter uncertainties and correlations. We do not find any noticeable correlation between any of the saturation parameters apart from $K$ and $E_{\rm sym}$, which show a moderate correlation (Pearson coefficient = 0.58; see appendix for further details). The positive $K$-$E_{\rm sym}$ correlation set by nuclear theory remains almost unchanged by astrophysical data. We do not observe any significant correlation between $E_{\rm sym}$ and $L$, similar to \textcite{Malik:2023mnx}. However, \textcite{Passarella:2025zqb} found a strong correlation between these two parameters, which can likely be attributed to their different coupling scheme for $\rho$-meson. Also, we observe that, as additional constraints are added sequentially, some credible regions significantly decrease in size. This sequential reduction results from the of the contraction of the posterior distributions, i.e., improved parameter estimation and reduced uncertainty as more information is incorporated into the analysis.


\subsection{Couplings}
\begin{figure*}[h]
    \centering
    \includegraphics[width=1\linewidth]{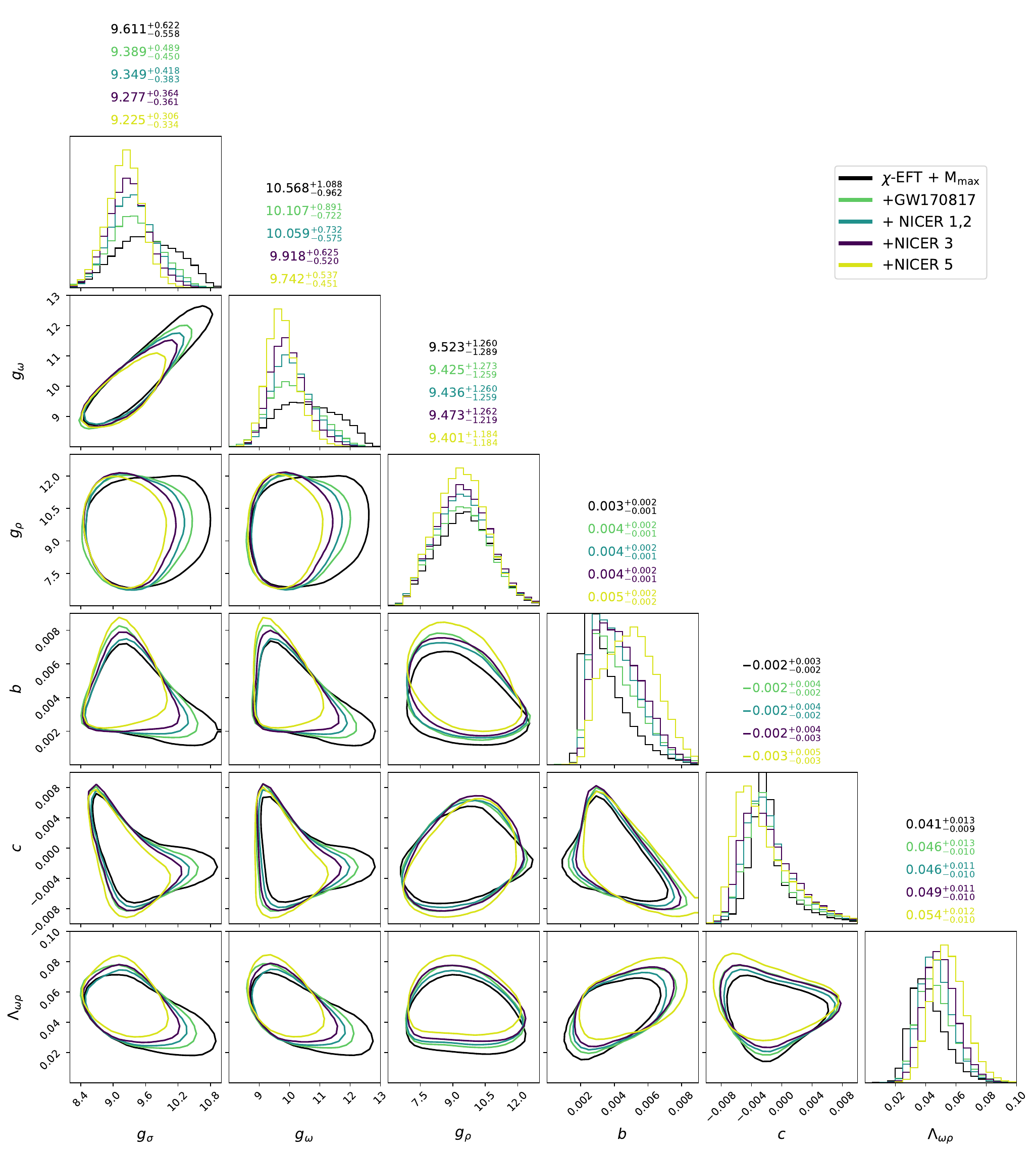}
    \caption{Same as Figure \ref{fig:corner_saturation} but now for coupling constants.}
    \label{fig:corner_couplings}
\end{figure*}    

\begin{table*}
    \centering
    \caption{Posterior 90\% CIs for the coupling parameters}
    \small
    \begin{tabular}{l c c c c c c}
        \hline\hline
        Parameters & \hspace{10pt} Unit & \hspace{10pt} $\chi$-EFT+ $\text{M}_\text{max}$  & \hspace{10pt} +GW170817 & \hspace{10pt} + NICER 1,2 & \hspace{20pt} + NICER 3 & \hspace{10pt} +NICER 5\\
        \hline

        $g_\sigma$ & \hspace{10pt} - 
        & \hspace{10pt} $9.611_{-0.875}^{+0.922}$  
        & \hspace{10pt} $9.389_{-0.736}^{+0.801}$ 
        & \hspace{10pt} $9.349_{-0.618}^{+0.714}$ 
        & \hspace{10pt} $9.277_{-0.588}^{+0.647}$ 
        & \hspace{10pt} $9.225_{-0.560}^{+0.525}$ \\

        $g_\omega$ & \hspace{10pt} - 
        & \hspace{10pt} $10.568_{-1.429}^{+1.595}$ 
        & \hspace{10pt} $10.107_{-1.113}^{+1.436}$ 
        & \hspace{10pt} $10.059_{-0.893}^{+1.278}$ 
        & \hspace{10pt} $9.918_{-0.819}^{+1.146}$ 
        & \hspace{10pt} $9.742_{-0.733}^{+0.952}$ \\

        $g_\rho$ & \hspace{10pt} - 
        & \hspace{10pt} $9.523_{-2.030}^{+2.027}$ 
        & \hspace{10pt} $9.425_{-1.936}^{+2.108}$ 
        & \hspace{10pt} $9.436_{-1.952}^{+2.083}$ 
        & \hspace{10pt} $9.473_{-1.914}^{+2.107}$ 
        & \hspace{10pt} $9.401_{-1.847}^{+2.021}$ \\

        $b$ & \hspace{10pt} - 
        & \hspace{10pt} $0.003_{-0.001}^{+0.003}$ 
        & \hspace{10pt} $0.004_{-0.002}^{+0.003}$ 
        & \hspace{10pt} $0.004_{-0.002}^{+0.003}$ 
        & \hspace{10pt} $0.004_{-0.002}^{+0.003}$ 
        & \hspace{10pt} $0.005_{-0.002}^{+0.003}$ \\

        $c$ & \hspace{10pt} - 
        & \hspace{10pt} $-0.002_{-0.003}^{+0.008}$ 
        & \hspace{10pt} $-0.002_{-0.004}^{+0.009}$ 
        & \hspace{10pt} $-0.002_{-0.004}^{+0.008}$ 
        & \hspace{10pt} $-0.002_{-0.004}^{+0.009}$ 
        & \hspace{10pt} $-0.003_{-0.004}^{+0.009}$ \\

        $\Lambda_{\omega\rho}$ & \hspace{10pt} - 
        & \hspace{10pt} $0.041_{-0.014}^{+0.023}$ 
        & \hspace{10pt} $0.046_{-0.016}^{+0.023}$ 
        & \hspace{10pt} $0.046_{-0.015}^{+0.020}$ 
        & \hspace{10pt} $0.049_{-0.016}^{+0.020}$ 
        & \hspace{10pt} $0.054_{-0.017}^{+0.021}$ \\

        \hline\hline
    \end{tabular}
    \label{tab:posterior_ranges_couplings_90CI}
\end{table*}

The coupling constants \(g_\sigma\), \(g_\omega\), \(g_\rho\), \(b\), \(c\), and $\Lambda_{\omega \rho}$ in the RMF theory represent the strengths of interactions between nucleons via meson exchanges and interactions between mesons. \autoref{fig:corner_couplings} shows the posterior distributions of the coupling constants and their correlations and \autoref{tab:posterior_ranges_couplings_90CI} records the corresponding 90\% CIs. We begin by examining the one‑dimensional marginalized distributions.

The trends observed in the coupling constants reflect how the RMF model adjusts to fit both nuclear theory and astrophysical observations. The median value of the scalar coupling $g_\sigma$ decreases  from 9.611 (baseline) to 9.225 (NICER-5), with a steady reduction in uncertainty, where GW170817 and NICER-5 play the most significant role.  The coupling $g_\omega$, which provides the dominant repulsive interaction, also shows the same sequential reduction.
 For $g_\rho$ the baseline constraint yields a very symmetric distribution 
 which remains largely unchanged on the application of further astrophysical data.
The small, stable values of \(b\) and \(c\) highlight the role of nonlinear self-interactions of $\sigma$-mesons in fine-tuning the EoS. These are already well constrained by nuclear theory and are not strongly impacted by astrophysical measurements.

The increase in $\Lambda_{\omega \rho}$ is particularly noteworthy, as it suggests that the interaction between the \(\omega\)-meson and \(\rho\)-meson which controls the density dependence of the symmetry energy becomes more significant when astrophysical data are included.  
It shows a clear upward trend, from 0.041 at baseline to 0.054 with NICER-5, along with steadily narrowing uncertainties. The sensitivity of $\Lambda_{\omega \rho}$ to these constraints highlights the growing role of NICER data in shaping the isovector channel’s behavior at supranuclear densities.

The joint posterior distributions of the couplings reveal several meaningful parameter correlations, reflecting how different sectors of the model adjust to simultaneously satisfy nuclear theory and astrophysical observations. Most prominent is the very strong positive correlation between the scalar and vector couplings, $g_\sigma$ and $g_\omega$. 
This correlation is already strong at baseline and becomes more compact, particularly under GW170817 and NICER-5, indicating that an increase in scalar attraction must be compensated by enhanced vector repulsion to satisfy the saturation properties and maximum mass constraints. This result is also consistent with what is found in literature across models and constraint sets\cite{Parmar:2026bmm,Huang:2023grj,Malik:2023mnx}.

Within the scalar sector, $g_\sigma$ shows a moderate negative correlation with the nonlinear scalar coupling $b$ at the baseline, which gradually weakens as observational constraints are added. A similar trend is observed between $g_\omega$ and $b$. In contrast, neither $g_\sigma$ nor $g_\omega$ exhibits meaningful correlation with the quartic scalar coupling $c$ across datasets, though a slight negative trend begins to emerge between $g_\sigma$ and $c$ with NICER-5. Both $g_\sigma$ and $g_\omega$ also display weak but growing negative correlations with the mixed isovector coupling $\Lambda_{\omega \rho}$, becoming moderately anti-correlated by NICER-5. This reflects a trade-off between the strength of isoscalar interactions and the density dependence of the symmetry energy, especially under tight mass-radius constraints of NICER 5. Among the nonlinear couplings, $b$ and $c$ are consistently negatively correlated, but this structure remains stable across constraints, with only the sampled area shifting. This behaviour of $b$ and $c$ is in agreement with recent studies \cite{Parmar:2026bmm,Huang:2023grj,Malik:2023mnx} even though they have an additional self interaction term for $\omega$-meson. Meanwhile, $b$ and $\Lambda_{\omega \rho}$ maintain a moderate positive correlation that also persists under constraint tightening. Overall, our two dimensional joint distributions for the couplings match well with \textcite{Malik:2023mnx} despite their different model and constraint set.

\subsection{Equation of state}
Next, we present plots for various physical quantities and observables. To generate these plots, we follow the procedure outlined below. First, we use the posterior samples to generate the corresponding EoSs and mass–radius (MR) relations. Then we fix a discrete set of mass values and, for each mass point, collect the corresponding radii from all the generated MR curves. This results in a distribution of radii at each mass. From these distributions, we compute the 90\% CI. Repeating this process across all mass points yields a band representing the 90\% credible region in the MR plane. The same method is applied across all successive constraints to generate the final diagram shown in \autoref{fig:mass_radius_td_plot}. We adopt a similar procedure to generate other similar plots.
\begin{figure}
    \centering
    \includegraphics[width=1\linewidth]{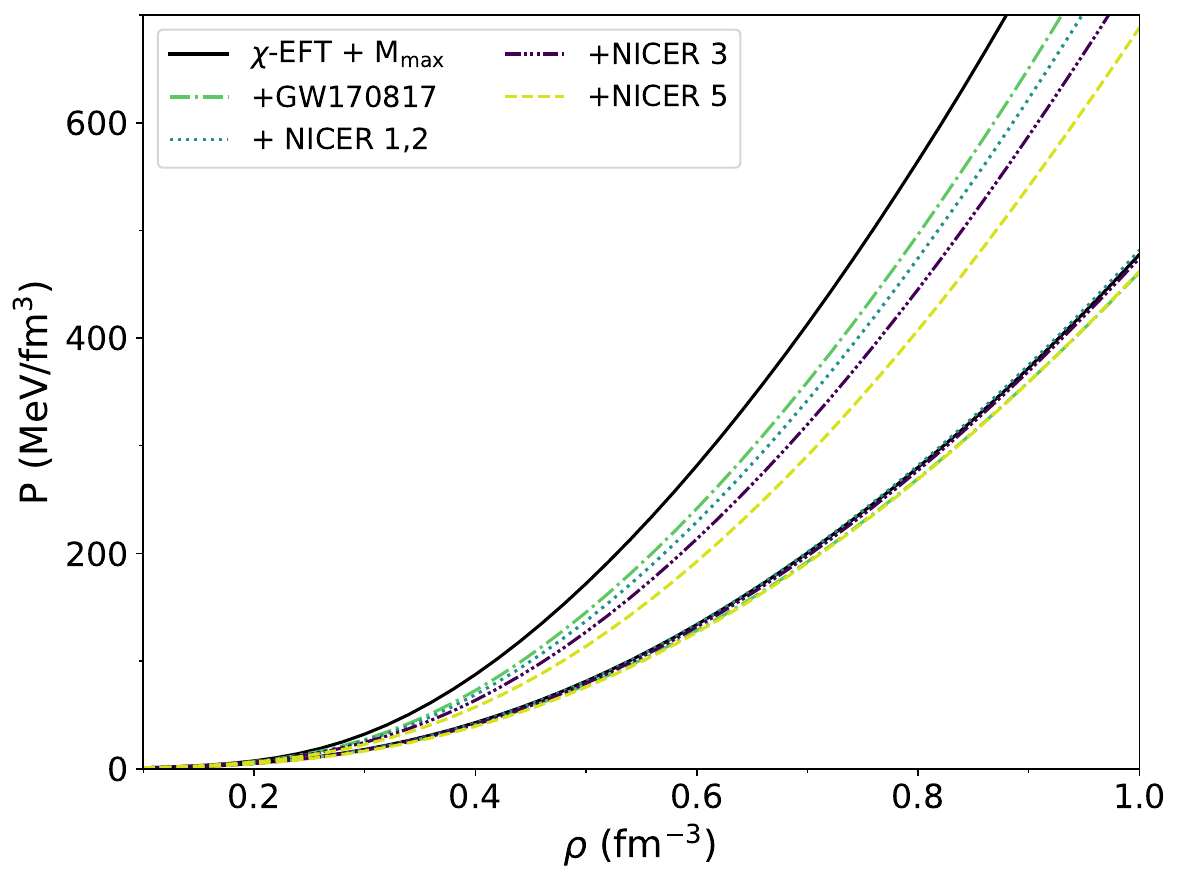}
    \caption{(a) Pressure vs baryon number density of matter in $\beta-$equilibrium. The different colored pairs of curves denote the 90 \% credible‐interval boundaries of the posterior samples as constraints are successively imposed in the order as shown in the legend.}
    \label{fig:presssure_number_density}
\end{figure}
Let's now analyze the individual figures one by one. A detailed examination of \autoref{fig:presssure_number_density} reveals the progressive impact of successive astrophysical constraints on the pressure–number density relation. The inclusion of GW170817 significantly narrows the 90\% credible band, notably shifting its upper boundary to lower pressure values compared to the $\chi$-EFT+$M_{\rm max}$ baseline. It indicates that GW170817 favors relatively softer equations of state (EoSs). This finding is consistent with previous studies\cite{Abbott:2018exr,Essick:2019ldf,Biswas:2020puz}. The addition of NICER 1 and 2 data further tightens the band, and the inclusion of NICER 3 reduces the area enclosed by the 90\% credible contours by approximately one-third relative to the baseline. The most substantial constraint comes from NICER 5, which shifts the upper boundary even closer to the lower one, effectively eliminating nearly half of the previously allowed region. Overall, the sequential application of astrophysical observations systematically narrows the 90\% credible region, increasingly favoring softer EoSs.

\subsection{Neutron Star Properties}

\begin{figure*}  
    \centering
    \subfigure[]{  
        \includegraphics[width=1\columnwidth]{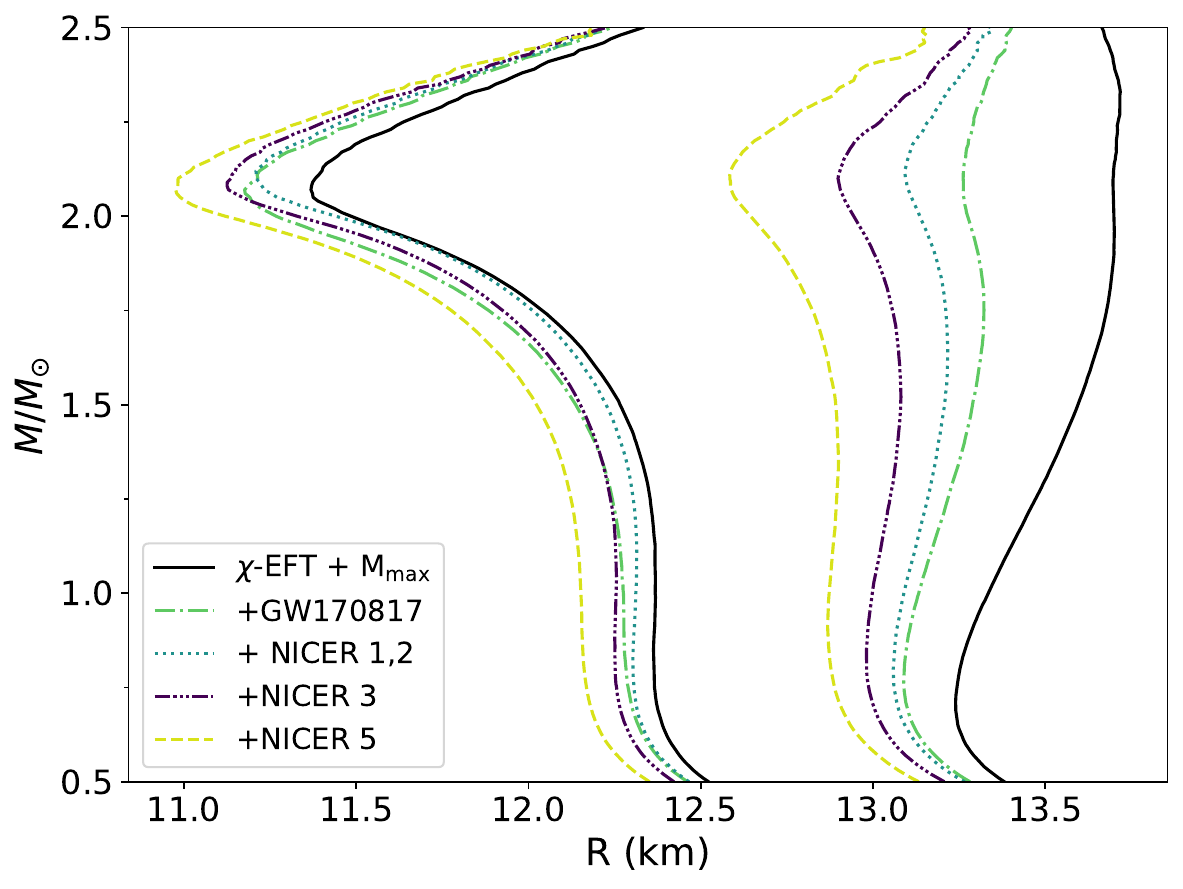}
        \label{fig:mass_radius_plot}
    }
    \hfill
    \subfigure[]{  
        \includegraphics[width=1\columnwidth]{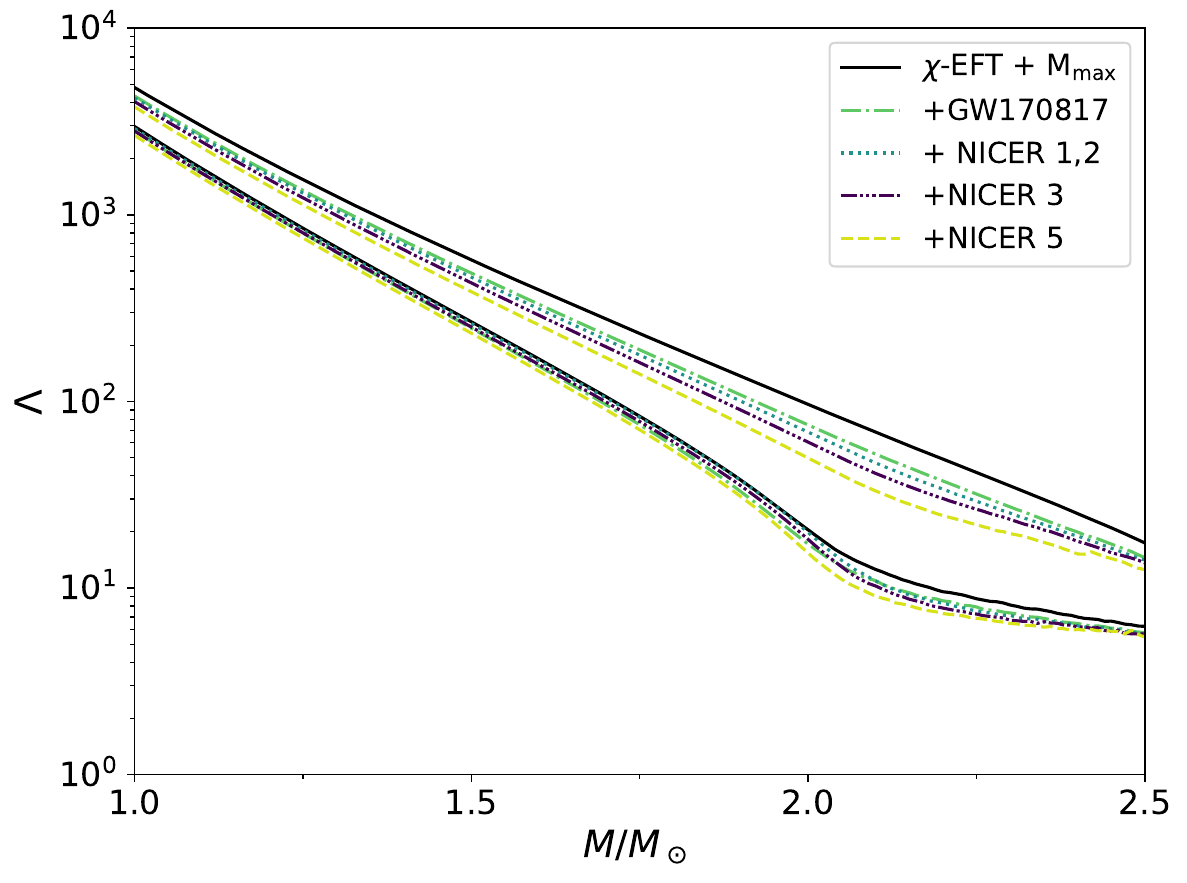}
        \label{fig:mass_tidal deformability_plot}
    }
    \caption{The left panel (a) shows the 90\% mass-radius contours and the right panel (b) mass vs tidal deformability contours for neutron stars  corresponding to the posterior samples. }
    \label{fig:mass_radius_td_plot}
\end{figure*}

Next, we turn to Figure \autoref{fig:mass_radius_plot}, which presents the MR diagram. The widest contour (corresponding to the baseline constraint) spans radii from approximately 12.3 to 13.5 km at $M = 1.4\,M_\odot$.
The inclusion of GW170817 removes a substantial portion of the MR space, particularly shifting the right boundary toward smaller radii and disfavoring larger radius values. This indicates that GW170817, when combined with $\chi$-EFT and the maximum mass constraint, favors softer EoSs. Adding the first two NICER mass–radius posteriors further trims the allowed region. The inclusion of NICER 3 tightens the contour even more and shifts it slightly leftward again, reinforcing the preference for softer EoSs. Finally, with the addition of NICER 5, the entire 90\% credible region shifts further to the left, significantly narrowing the band and strongly favoring the softest EoS models considered. The dramatic reduction in the final contour relative to the baseline $\chi$-EFT+$M_{\rm max}$ region underscores the power of multimessenger observations in determining neutron-star properties.

In Figure \autoref{fig:mass_tidal deformability_plot}, we present the mass–tidal deformability ($M$–$\Lambda$) relationship for neutron stars. 
Just like before, we observe a progressive narrowing of the tidal deformability bands with each added constraint. Initially, the deformability at each mass exhibits considerable uncertainty. The inclusion of the GW170817 measurement significantly reduces the 90\% CI, reflecting a strong preference for lower tidal deformability values—indicative of a softer EoS. Subsequent incorporation of the NICER 1 and 2 observations further shrinks the 90\% region. The addition of NICER 3 and 5 has a comparable effect to that of GW170817, tightening the bands even more and reinforcing the trend toward a softer EoS.
At higher masses, the tidal deformability bands begin to widen slightly. It suggests that current nuclear and astrophysical results are less effective at constraining $\Lambda$ for massive stars. Nonetheless, the overall uncertainty is still significantly reduced compared to the initial, unconstrained scenario.

\subsection{Composition of neutron star matter}
\begin{figure*}  
    \centering
    \subfigure[]{  
        \includegraphics[width=1\columnwidth]{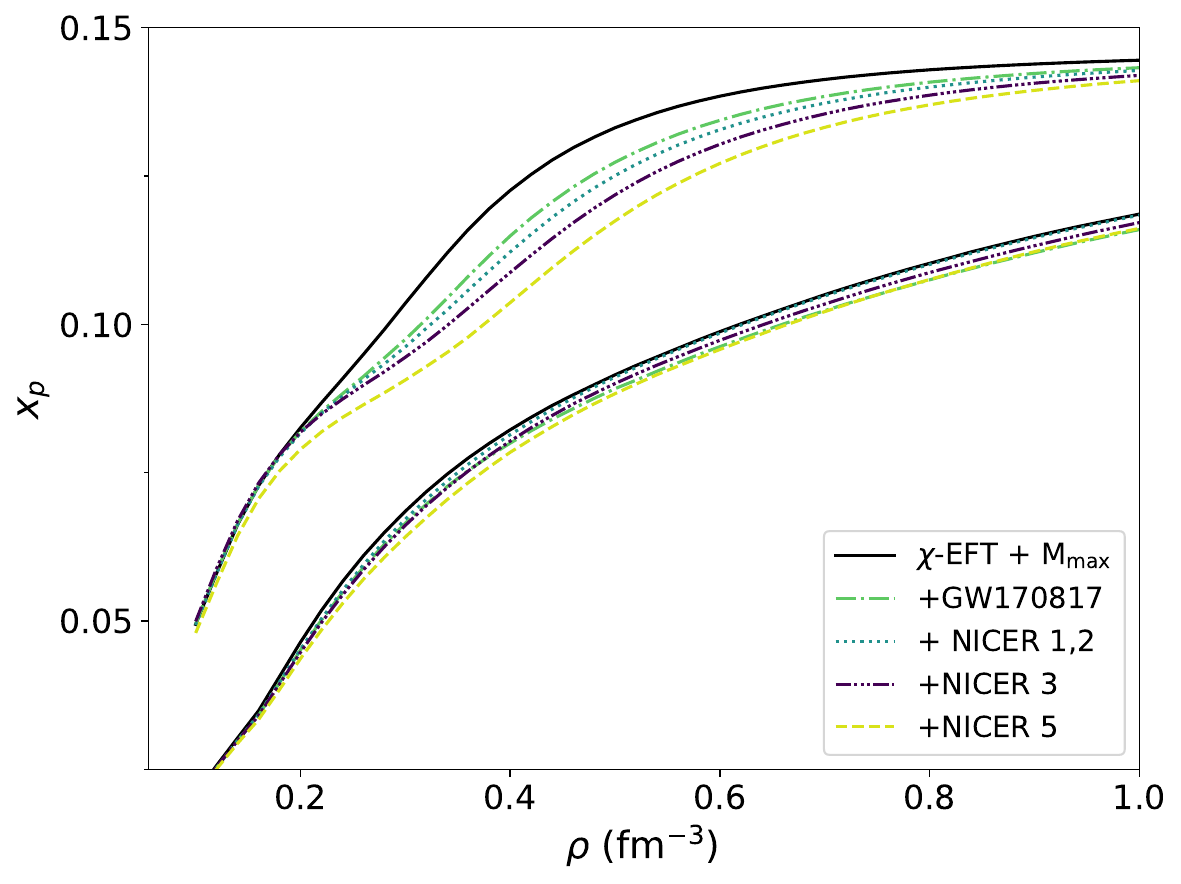}
        \label{fig:pfraction_number_density_plot}
    }
    \hfill
    \subfigure[]{  
        \includegraphics[width=1\columnwidth]{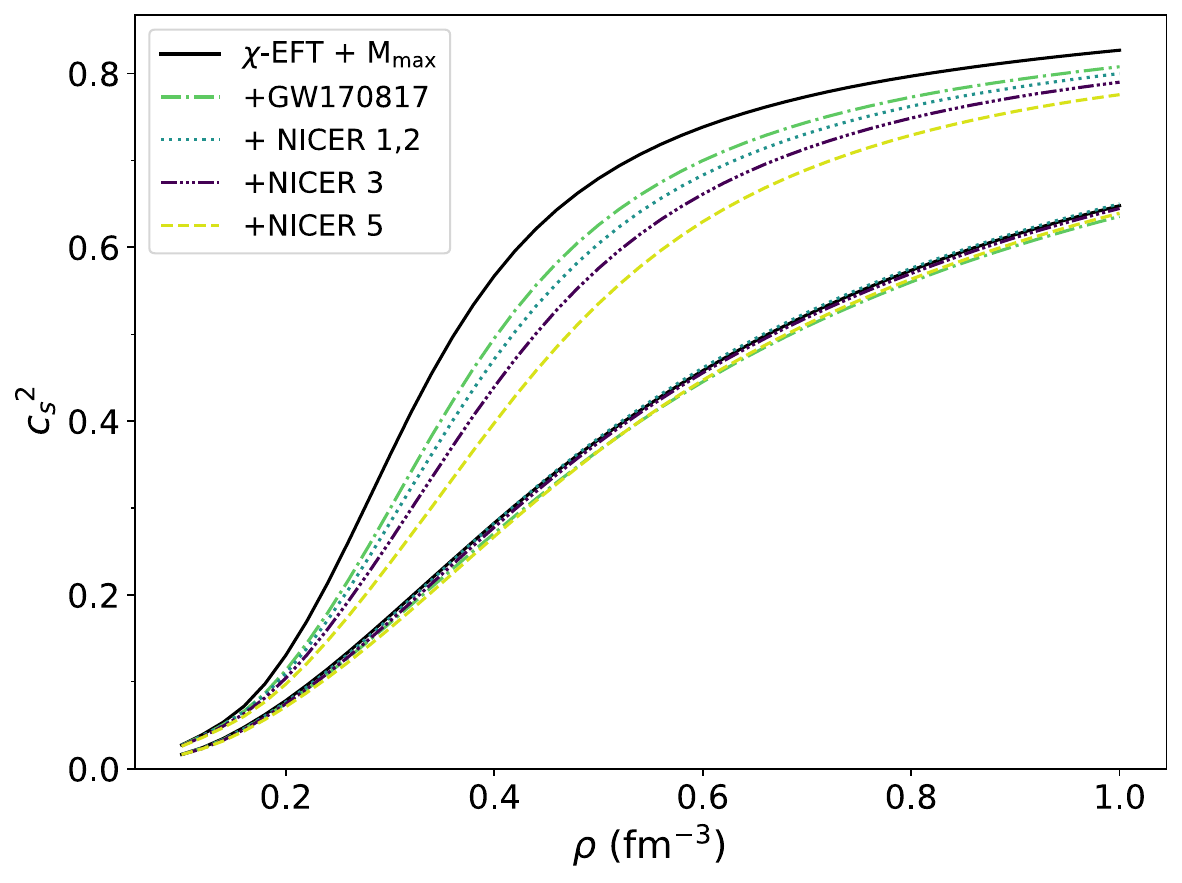}
        \label{fig:csound_number_density_plot}
    }
    \caption{Same  as Fig. 3 but for (a) Proton fraction vs Number density, and (b) Sound Speed squared vs number density.}
    \label{fig:pfraction_csound_plot}
\end{figure*}

We show the $90\%$ CIs for proton fraction $x_p$ as a function of number density in \autoref{fig:pfraction_number_density_plot}. Across all datasets, $x_p$ increases monotonically from nearly zero at  $\rho \approx 0.05\,\mathrm{fm}^{-3}$, with the upper boundary of the CI band reaching just below  0.15 and the lower boundary slightly above 0.10 by $\rho = 1.0\,\mathrm{fm}^{-3}$. This result is consistent with \textcite{Malik:2023mnx}, \textcite{Huang:2023grj} and the non-linear model of \textcite{Cartaxo:2025jpi}, despite their additional $\omega$ self interaction term. However, some studies using density-dependent couplings found support for much higher proton fractions \cite{Char:2023fue,Scurto:2024ekq,Scurto:2025bqr}. Up to $\rho\approx0.2\,\mathrm{fm}^{-3}$, all five $90\%$ credible bands are nearly indistinguishable. Above $\rho\approx0.2$ $\mathrm{fm}^{-3}$, all the bands begin to widen, with the baseline band exhibiting the greatest broadening. Incorporating the GW170817 constraint narrows this band and shifts it slightly downward, indicating a preference for a lower proton fraction. While the inclusion of NICER 1 and 2 data has little impact, NICER 3 and 5 significantly tighten the credible bands, highlighting the complementary role of mass-radius measurements in constraining proton fraction at these densities. This information is crucial for determining the threshold of the direct Urca process, which governs the cooling of the neutron stars and their bulk viscosity. At higher densities ($\rho\sim0.9$ $\mathrm{fm}^{-3}$), the bands begin to converge again, with little variation between the different datasets.

Since we are working within a relativistic mean-field framework, the theory is inherently causal. It is confirmed in \autoref{fig:csound_number_density_plot}, where the sound speed squared never approaches unity (the speed of light squared). At low densities, $c_s^2$ remains small and does not vary much between EoSs, as indicated by the narrow 90\% CI. However, beyond  $\rho\sim0.2$ fm$^{-3}$, it increases rapidly and the uncertainty band broadens, reaching its maximum width near 0.5 fm$^{-3}$. Beyond this point, the $c_s^2$ grows more slowly, and the credible band narrows slightly. The impact of successive observational constraints follows a similar trend to that seen in previous figures. Notably, GW170817 and NICER 5 data impose the most stringent constraint, significantly narrowing the baseline band across the relevant density range.


\begin{figure}
    \centering
    \includegraphics[width=1\columnwidth]{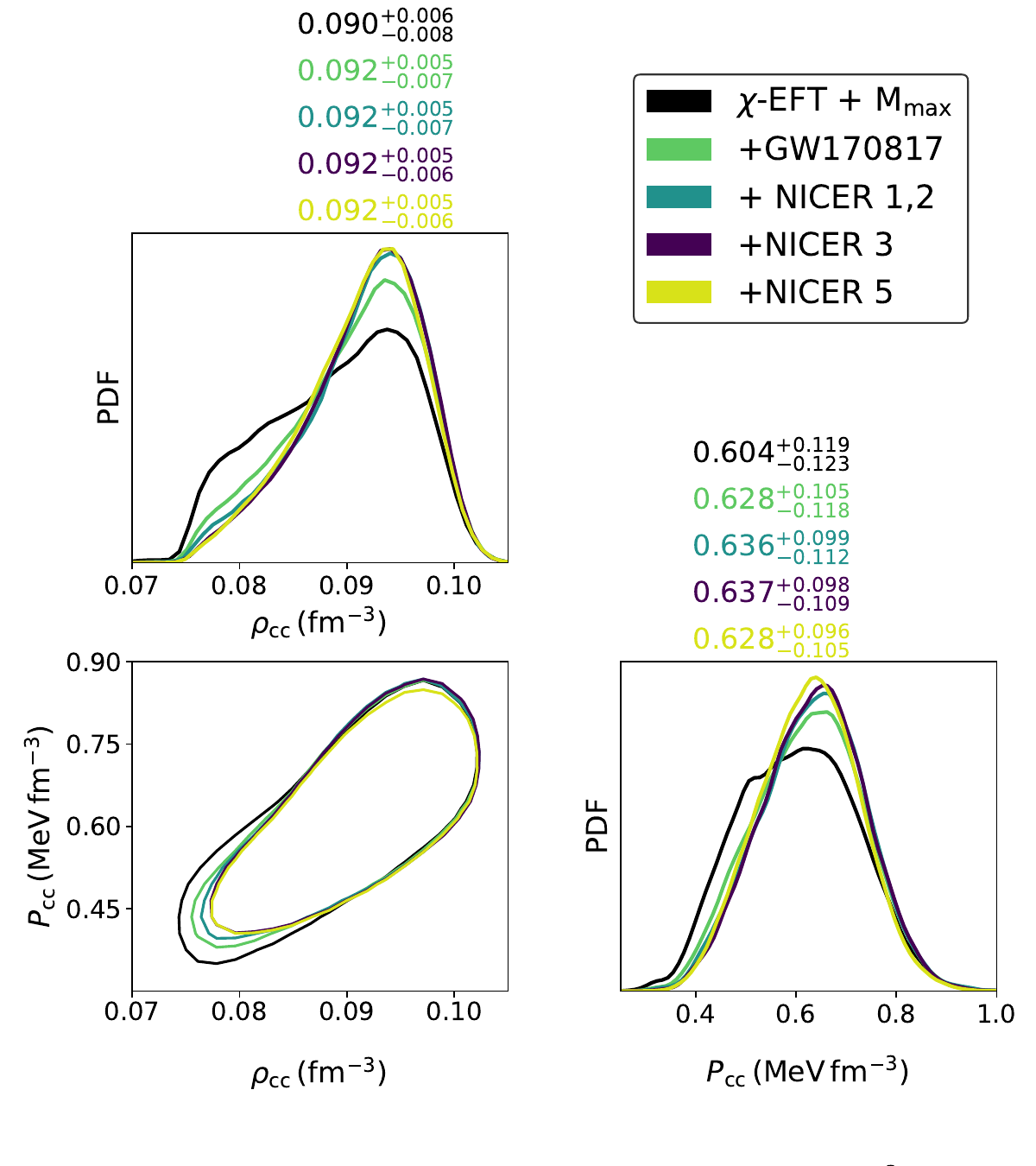}
    \caption{Corner plot depicting the posterior distributions of core-crust transition density and transition pressure. The diagonal panels display the one-dimensional marginalized distributions for each, while the off-diagonal panels show the 90\% contours for the two-dimensional joint probability density functions. The colored contours and distributions represent
    successive constraints applied in the order indicated by the legend. On top of each diagonal panels we also show the 68\% CI for each parameter in the order indicated by the legend.}
    \label{fig:transition_cc}
\end{figure}

In \autoref{fig:transition_cc}, we present a corner plot depicting the core–crust transition  density ($\rho_{cc}$) and pressure ($P_{cc}$) of the neutron star EoS, with successive observational constraints applied in the same manner as in previous figures. We observe that the transition density and pressure exhibit an increase in their median values following the inclusion of the GW170817 constraint, while also showing a slight decrease in the 68\% CI. On application of other constraints successively, the distribution of transition density remains largely unchanged. In contrast, the median pressure shows some movement,  increasing slightly from 0.628 to 0.637 MeV fm$^{-3}$, and coming back to 0.628 MeV fm$^{-3}$ upon the inclusion of NICER 5. Also, the 68\% CIs for transition pressure shrink continuously after the application of successive constraints. Additionally, the 2D plot shows a clear positive correlation between the transition density and transition pressure for all the cases. 

As discussed in \autoref{EoS-Section}, we determine the crust-core transition by finding the point at which the incompressibility at constant chemical potential,  $K_\mu$, becomes negative. Since $K_\mu$ depends on the incompressibility of symmetric nuclear matter ($K$) and symmetry energy expansion coefficients ($L$ and $K_{\rm sym}$), the transition point is also sensitive to these coefficients. We discard all combinations of these coefficients for which $K_\mu$ does not change sign before reaching the outer crust, and consequently the EoS derived from them. Evidently, this has a direct bearing on our overall results when constraining the EoS with astrophysical observations. Modifying the criterion for the crust-core transition point, for instance by including finite-size effects, could therefore alter which EoS survive this selection, and hence potentially affect our overall results. We plan to explore this in a future study. 

\subsection{The 1.4 M$_\odot$ and the Maximum Mass Neutron Star}
\begin{figure*}[t]
    \centering
    \begin{minipage}{0.5\textwidth}
        \centering
        \includegraphics[width=\linewidth]{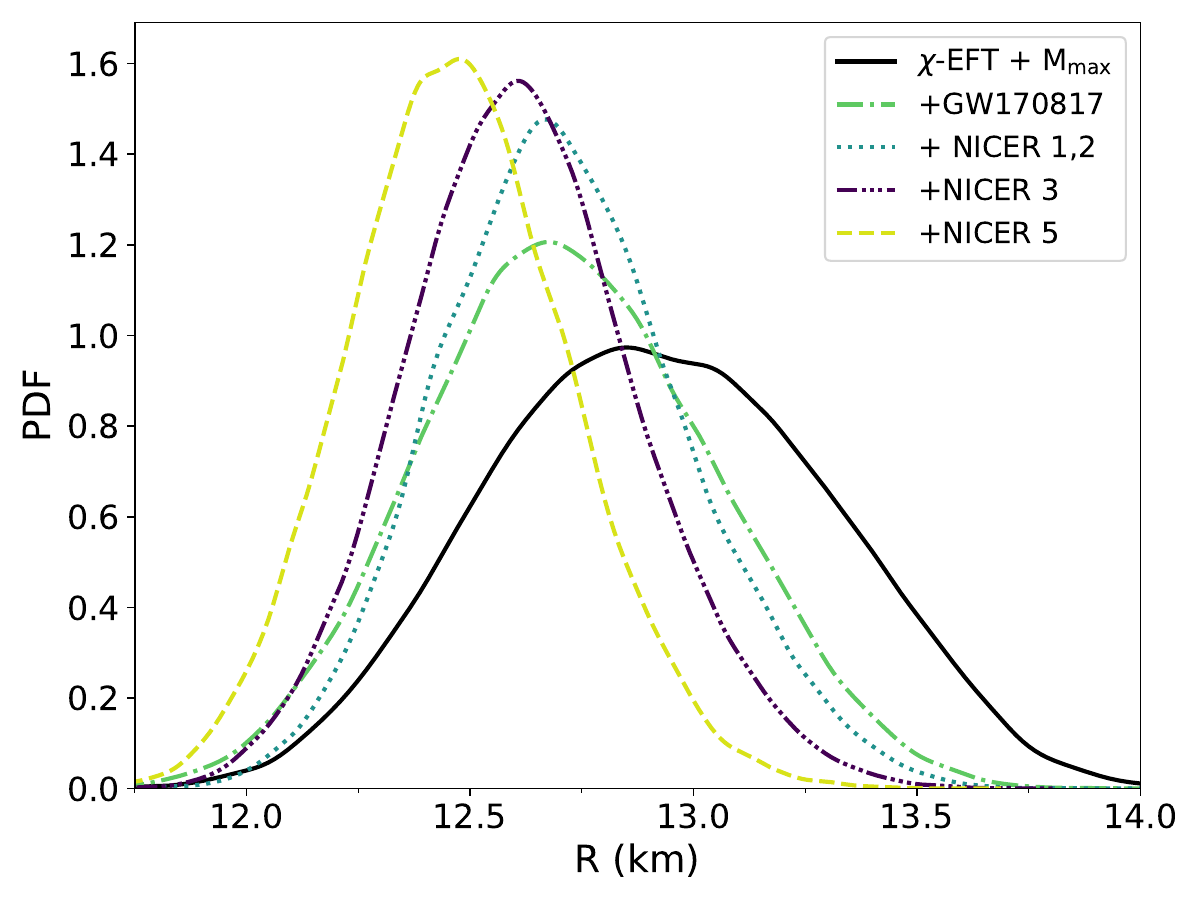}
        
    \end{minipage}%
    \hfill
    \begin{minipage}{0.5\textwidth}
        \centering
        \includegraphics[width=\linewidth]{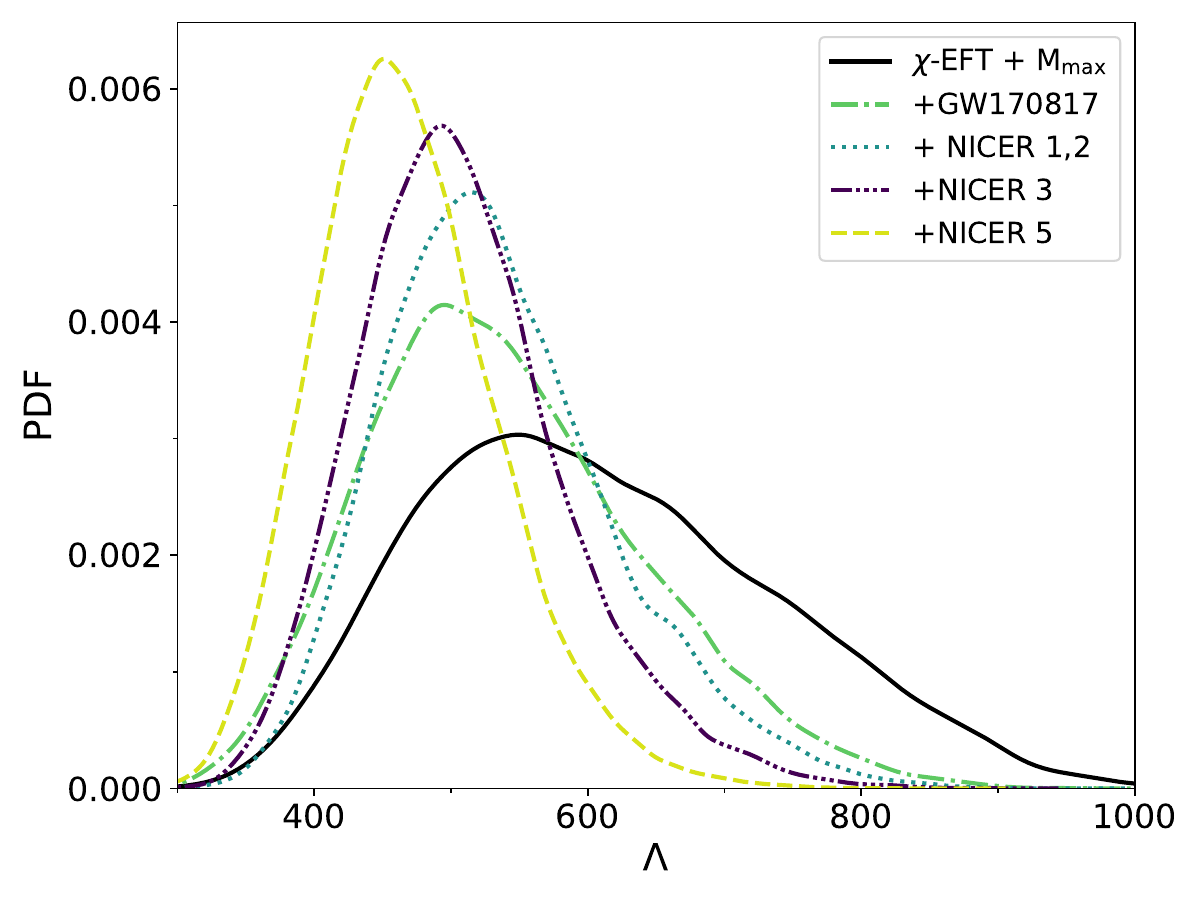}
        
    \end{minipage}

    \vspace{0.4cm}
    
    \begin{minipage}{0.5\textwidth}
        \centering
        \includegraphics[width=\linewidth]{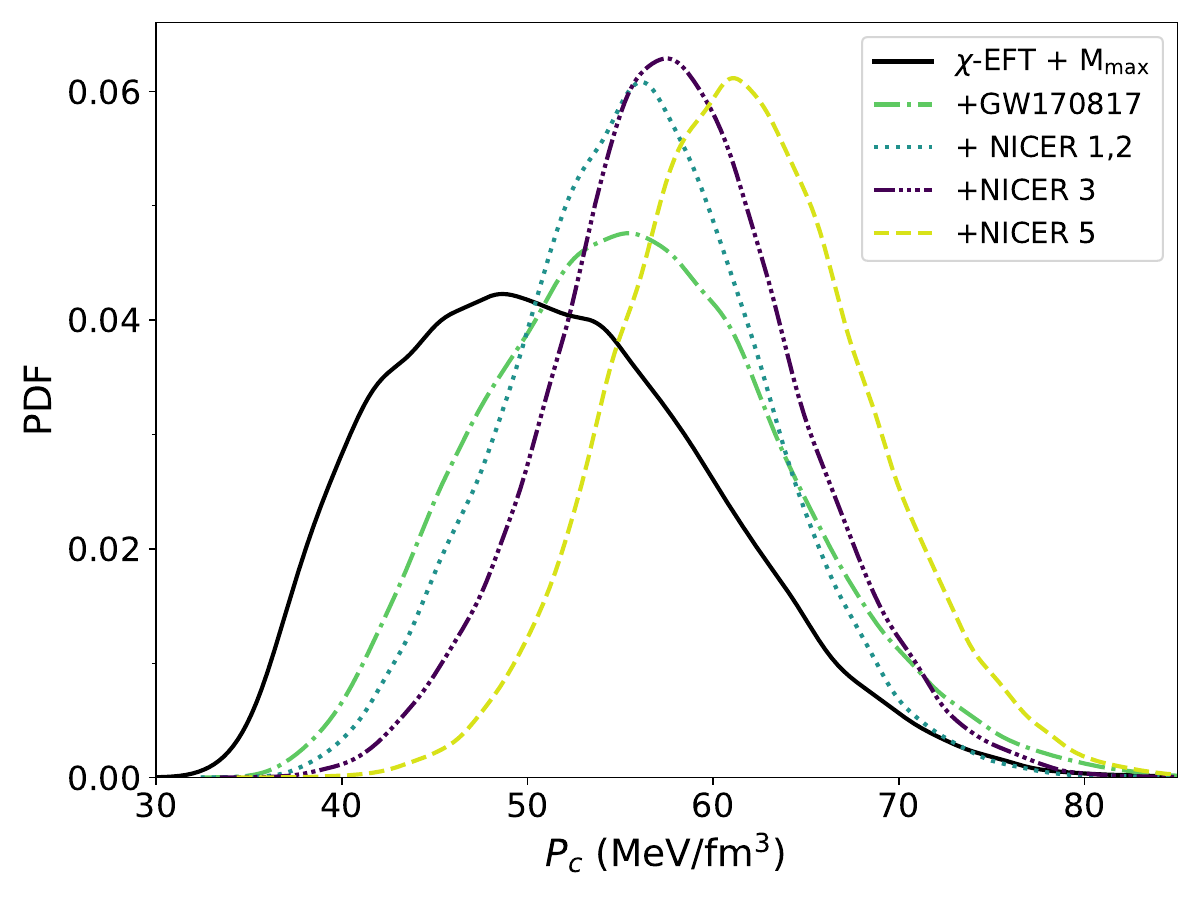}
        
    \end{minipage}
    \caption{The three panels above show the posterior probability distributions of radius (R), dimensionless tidal deformability $(\Lambda)$, and central pressure $(P_c)$ for a canonical 1.4 $M_\odot$ neutron star. The constraints are applied in the order as shown in the legend.}
    \label{fig:M1_4}
\end{figure*}

\begin{table*}
\caption{\label{tab:posterior_ranges_1p4_90CI}
90\% Posterior CIs for Key Physical Properties of a 1.4 $M_\odot$ Neutron Star}
\begin{ruledtabular}
\begin{tabular}{lcccccc}
Physical Quantity & Unit & $\chi$-EFT+ $M_\mathrm{max}$ & +GW170817 & + NICER 1,2 & + NICER 3 & +NICER 5 \\
\hline

$R$ & km 
& $12.917_{-0.604}^{+0.633}$ 
& $12.710_{-0.517}^{+0.555}$ 
& $12.697_{-0.429}^{+0.493}$ 
& $12.612_{-0.415}^{+0.460}$ 
& $12.471_{-0.394}^{+0.427}$ \\

$\Lambda$ & -- 
& $595.530_{-174.490}^{+245.164}$ 
& $528.943_{-136.799}^{+193.580}$ 
& $525.287_{-113.260}^{+166.415}$ 
& $501.368_{-104.944}^{+146.814}$ 
& $463.019_{-94.384}^{+124.112}$ \\


$P_c$ & MeV/fm$^3$ 
& $50.673_{-12.174}^{+15.649}$ 
& $55.648_{-12.301}^{+14.439}$ 
& $55.962_{-11.062}^{+11.315}$ 
& $57.911_{-10.642}^{+11.344}$ 
& $61.484_{-10.373}^{+11.553}$ \\

\end{tabular}
\end{ruledtabular}
\end{table*}

In \autoref{fig:M1_4}, we show the posterior distributions of radius, tidal deformability, and central pressure for  1.4$M_\odot$ neutron stars.
The median radius, based on the baseline analysis is found to be 12.92 km. Upon including the GW170817 constraint, this value drops to 12.71 km, indicating that GW170817 favors a softer EoS. A softer EoS has less pressure at a given density, leading to more compact stars (i.e., smaller radii for a given mass, $1.4\,M_\odot$ here). The inclusion of NICER 1 and 2 does not significantly alter the radius of a 1.4\,$M_\odot$ neutron star, while the median radius again drops to 12.61 km when NICER 3 is incorporated into the analysis. The addition of NICER 5 on top of the previous constraints results in a larger reduction in the median radius down to 12.47 km, once again indicating a strong preference for a softer EoS. Tidal deformability and central pressure consistently reflect this same behavior. The key feature of this figure is the reduction in uncertainties relative to the baseline scenario with the application of successive constraints, consistent with the trends seen in earlier figures. The median and CIs of the three quantities in \autoref{fig:M1_4} are listed in \autoref{tab:posterior_ranges_1p4_90CI}.


\begin{figure*} 
    \centering
    \subfigure[]{  
        \includegraphics[width=1\columnwidth]{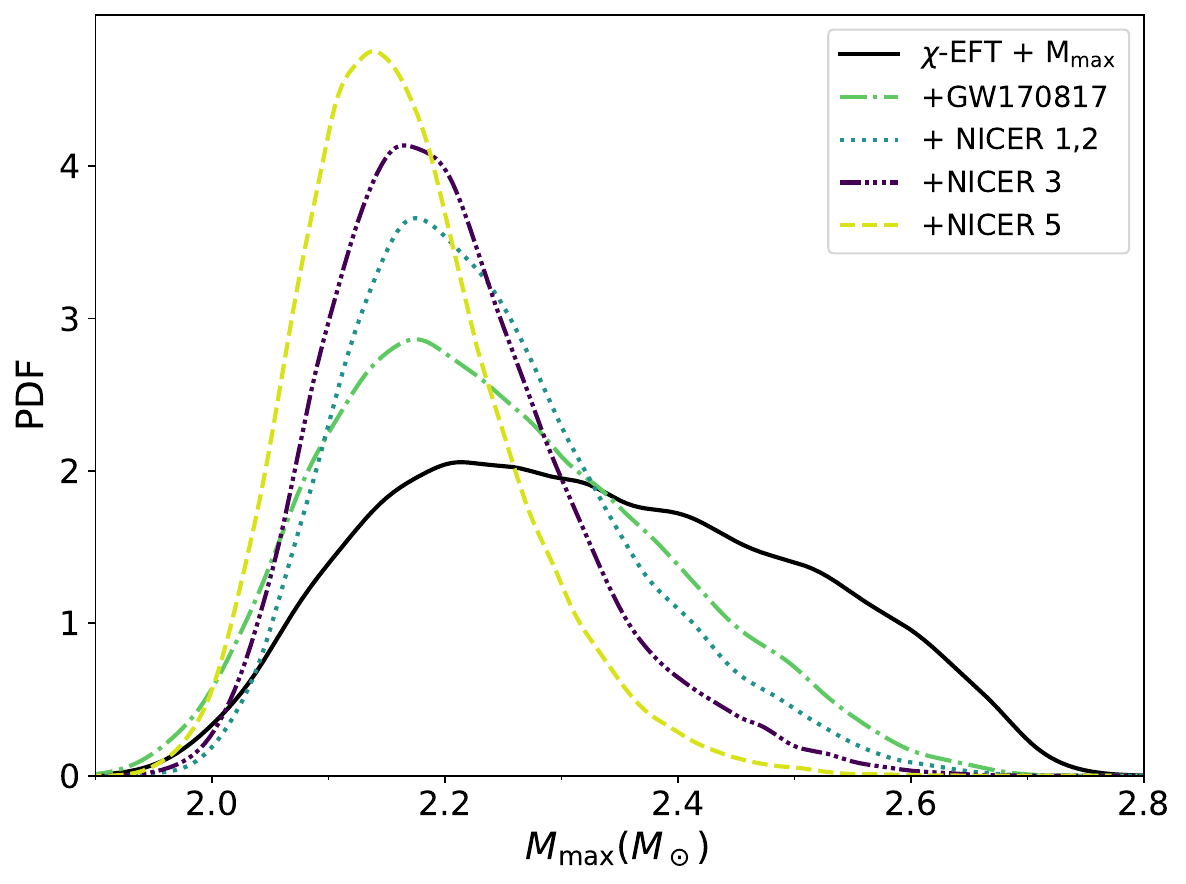}
        \label{fig:mmax_distribution}
    }
    \hfill
    \subfigure[]{  
        \includegraphics[width=1\columnwidth]{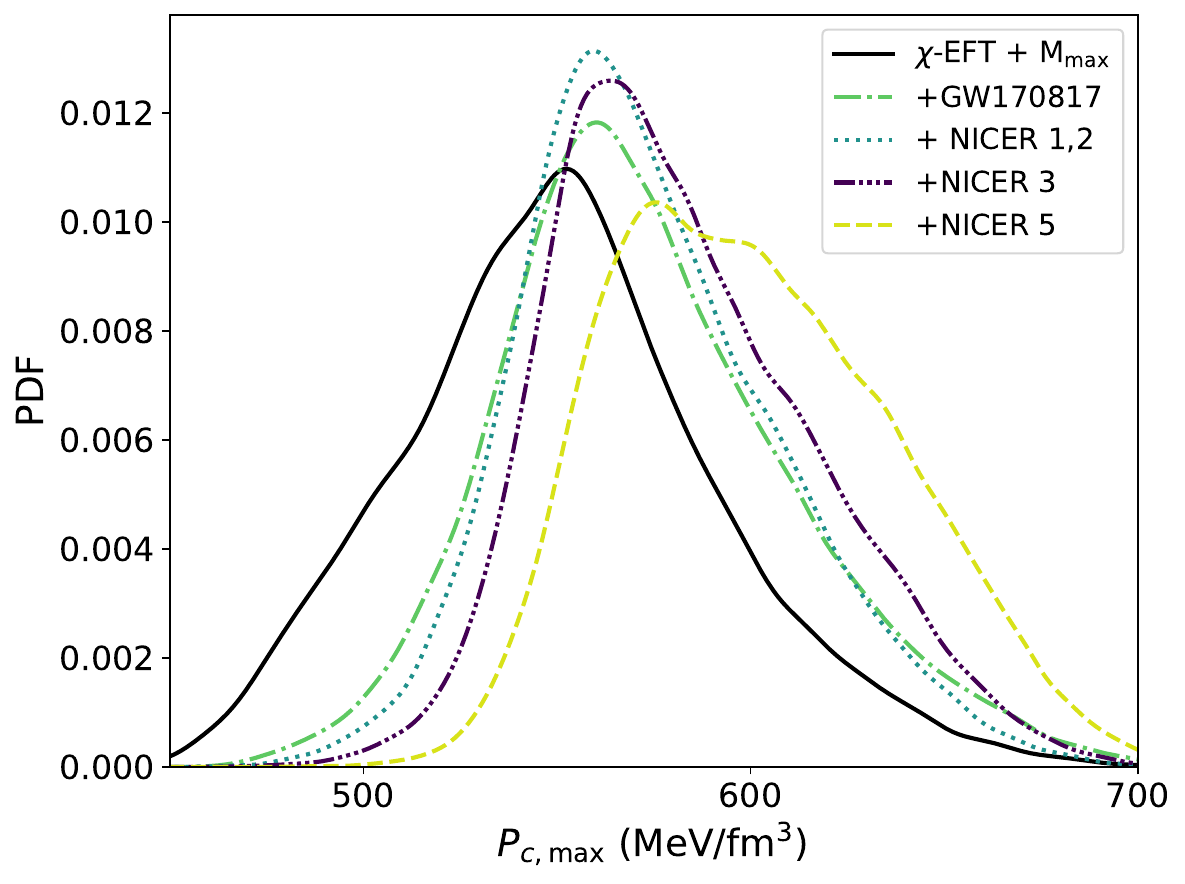}
        \label{fig:Pc_Mmax_distribution}
    }
    \caption{ One‐dimensional posterior probability densities for (a) the maximum neutron-star mass and (b) their associated central pressure. The constraints are applied in the order as shown in the legend.}
    \label{fig:mmax_Pcmmax_plot}
\end{figure*}

\begin{table*}
\caption{\label{tab:posterior_ranges_mmax_90CI}
90\% Posterior CIs for Key Physical Properties of Maximum Mass Neutron Star}
\begin{ruledtabular}
\begin{tabular}{lcccccc}
Physical Quantity & Unit & $\chi$-EFT+ $M_\mathrm{max}$ & +GW170817 & + NICER 1,2 & + NICER 3 & +NICER 5 \\
\hline

$M$ & $M_\odot$ 
  & $2.313_{-0.247}^{+0.306}$ 
  & $2.226_{-0.188}^{+0.271}$ 
  & $2.218_{-0.150}^{+0.238}$ 
  & $2.192_{-0.135}^{+0.212}$ 
  & $2.156_{-0.120}^{+0.175}$ \\

$R$ & km 
  & $11.457_{-0.778}^{+1.064}$ 
  & $11.163_{-0.611}^{+0.900}$ 
  & $11.142_{-0.483}^{+0.777}$ 
  & $11.047_{-0.454}^{+0.680}$ 
  & $10.893_{-0.423}^{+0.564}$ \\

$\mathcal{E}_c$ & MeV/fm$^3$ 
  & $1192.239_{-222.889}^{+217.099}$ 
  & $1267.472_{-213.038}^{+178.451}$ 
  & $1273.873_{-189.210}^{+138.062}$ 
  & $1298.371_{-172.821}^{+130.728}$ 
  & $1339.112_{-152.754}^{+126.116}$ \\

$P_c$ & MeV/fm$^3$ 
  & $550.043_{-64.436}^{+72.813}$ 
  & $568.864_{-52.653}^{+74.396}$ 
  & $569.650_{-46.610}^{+66.248}$ 
  & $577.645_{-44.144}^{+67.722}$ 
  & $597.241_{-50.157}^{+67.131}$ \\

\end{tabular}
\end{ruledtabular}
\end{table*}

In \autoref{fig:mmax_Pcmmax_plot}, we show the posterior distributions of mass and central pressure corresponding to the maximum mass neutron star. The inference about EoS stiffness drawn from the 1.4\,$M_\odot$ neutron star is consistently supported by the maximum mass neutron star plots as well. After incorporating all the constraints, we find the median maximum mass of the neutron star to be $2.156\,M_\odot$. In \autoref{tab:posterior_ranges_mmax_90CI} we show the posterior 90\% CIs of the same quantities which we have in \autoref{fig:mmax_Pcmmax_plot}, as well as radius and central energy density of the maximum mass neutron star.

When we compare our results with \textcite{Malik:2023mnx}, who did not use the GW170817 and NICER measurements in their likelihood, we find that our corresponding median values (column 3 of \autoref{tab:posterior_ranges_1p4_90CI} and \autoref{tab:posterior_ranges_mmax_90CI}) for the maximum mass, radius and tidal deformability of $1.4M_\odot$ are larger. Similar trends are also seen when comparing our results (column 6 of \autoref{tab:posterior_ranges_1p4_90CI} and \autoref{tab:posterior_ranges_mmax_90CI}) with \textcite{Cartaxo:2025jpi}, who used constraints up to NICER-3. This is possibly because they included an $\omega$ self-interaction term in their Lagrangian which softens the EoS, making the star more compact. We also find considerable overlap between our  distributions of radius and tidal deformability of a $1.4M_\odot$ star with the corresponding distributions of \textcite{Zhu:2022ibs}, who used the same Lagrangian as ours but did not apply the $\chi-$EFT constraint. The effect of NICER-3 and NICER-5 in softening the EoS that we observe here is consistent with other recent studies  (\cite{Saha:2025don,Biswas:2025ivu,Biswas:2024hja}) that adopted different approaches.

\subsection{Neutron Skin Thickness}
\begin{figure*}[ht]
    \centering
    \begin{minipage}{0.5\textwidth}
        \centering
        \includegraphics[width=\linewidth]{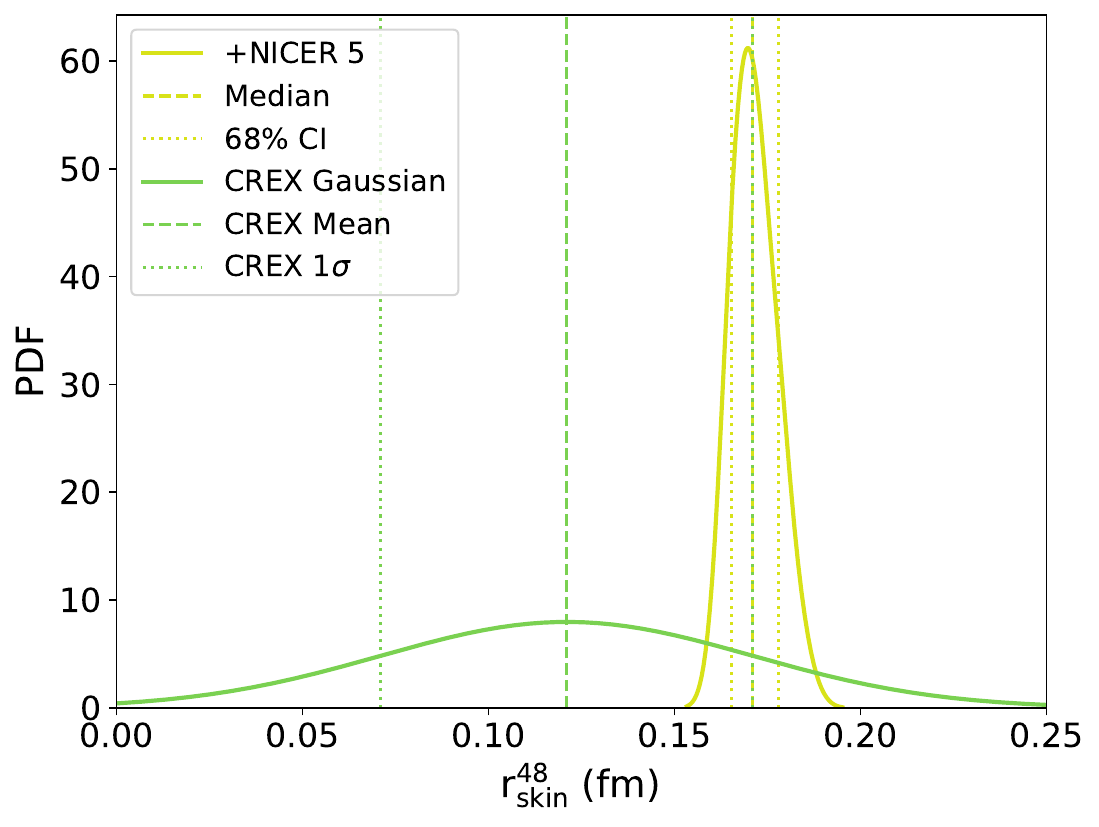}
        
    \end{minipage}%
    \hfill
    \begin{minipage}{0.5\textwidth}
        \centering
        \includegraphics[width=\linewidth]{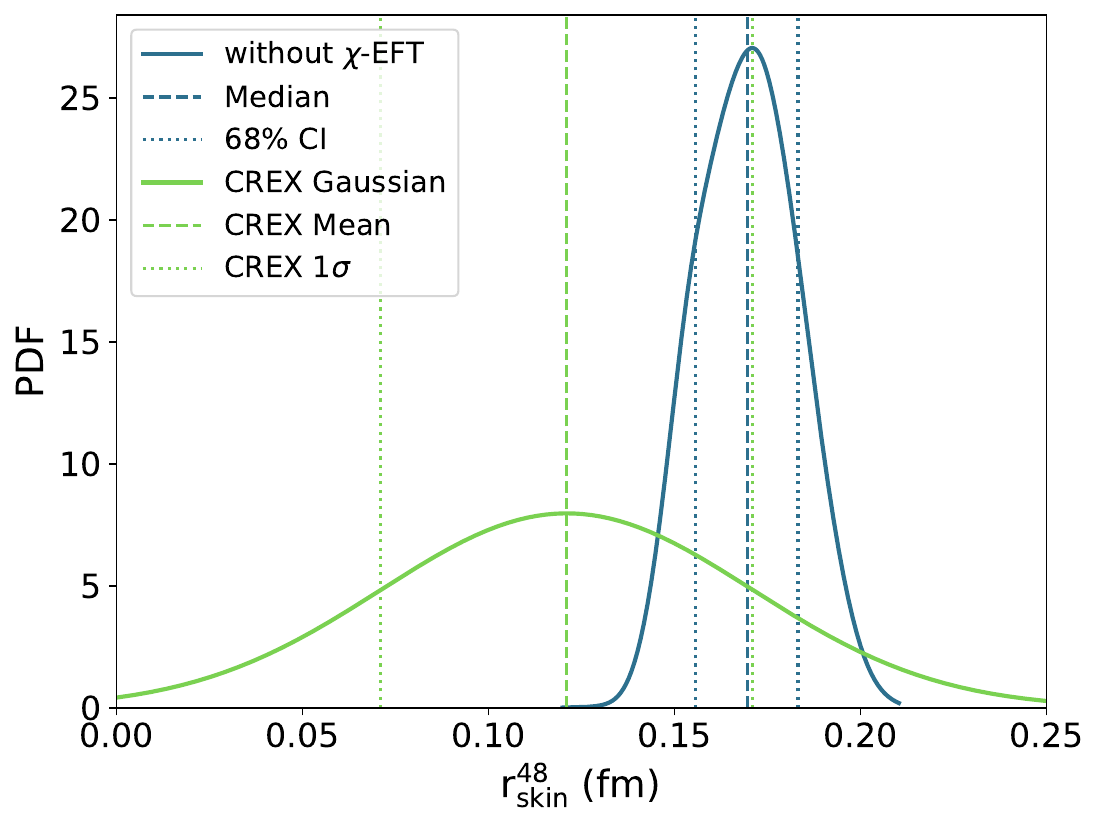}
        
    \end{minipage}

    \vspace{0.4cm}

    \begin{minipage}{0.5\textwidth}
        \centering
        \includegraphics[width=\linewidth]{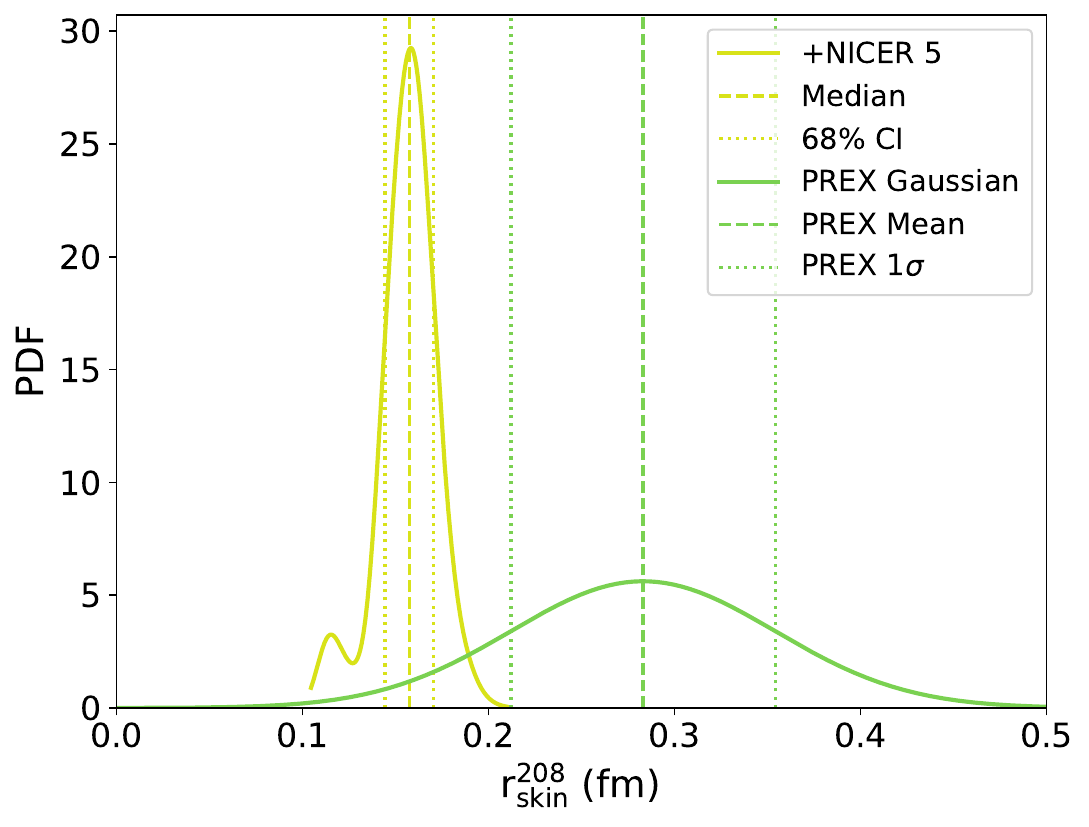}
        
    \end{minipage}%
    \hfill
    \begin{minipage}{0.5\textwidth}
        \centering
        \includegraphics[width=\linewidth]{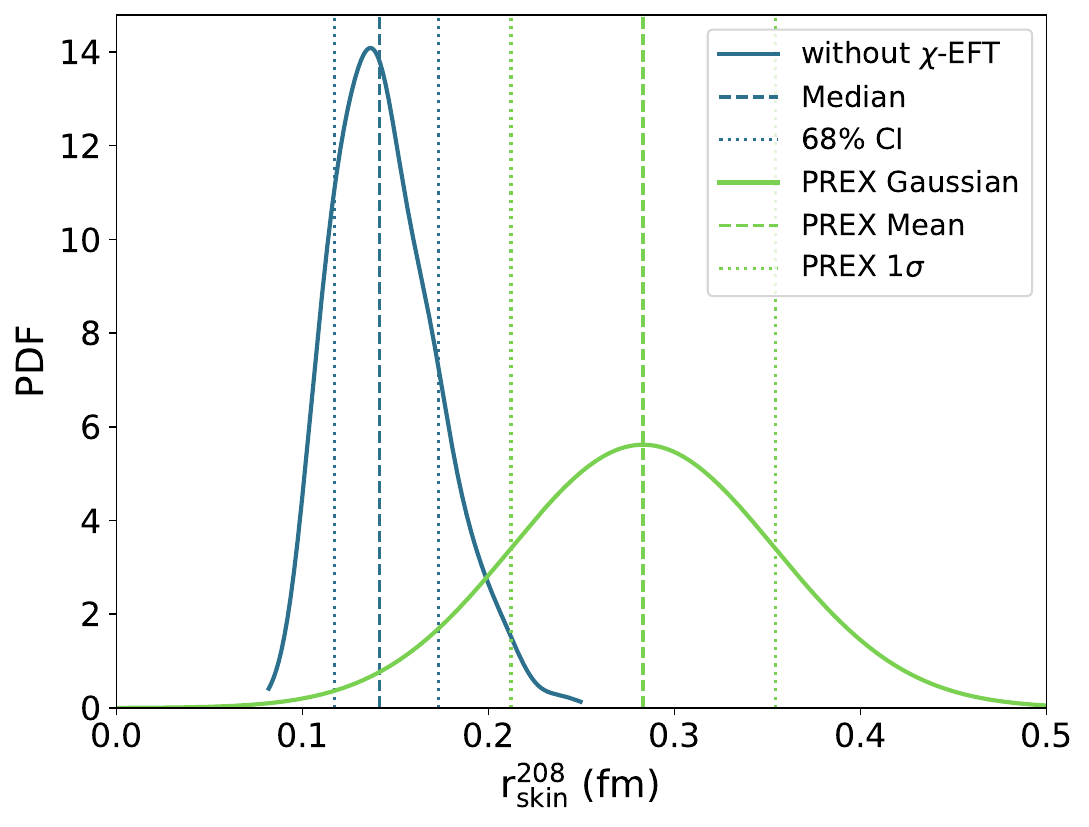}
        
    \end{minipage}

    \caption{\rkb{Neutron skin thickness distributions for $^{48}\mathrm{Ca}$ (top) and $^{208}\mathrm{Pb}$ (bottom) under different constraint sets. Left panels show skin distributions corresponding to our final constraint set (yellow), while right panels show skin distributions corresponding to all our constraints except $\chi-$EFT (steel blue), with CREX and PREX measurements shown in green.}}
    \label{neutron_skin_plot}
\end{figure*}
The neutron skin thickness ($\rm r_\mathrm{skin}$), defined as the difference between the neutron and proton root-mean-square radii ($\rm r_\mathrm{skin} = \langle r_n^2 \rangle^{1/2} - \langle r_p^2 \rangle^{1/2} $), is sensitive to the density dependence of symmetry energy especially its slope $L$, which governs the behavior of neutron-rich matter in both finite nuclei and neutron stars. We use our equally weighted posterior samples of the saturation properties to generate the corresponding sets of couplings. We then use these couplings to calculate $\rm r_\mathrm{skin}$ for Calcium-48 ($^{48}\mathrm{Ca}$) and Lead-208 ($^{208}\mathrm{Pb}$). This procedure is carried out for two constraint sets: one with all constraints (denoted as +NICER 5), and another with all constraints except $\chi$-EFT (without $\chi$-EFT). Note that, the code that we use to calculate $\rm r_\mathrm{skin}$ of nuclei, does not converge for all the sets of couplings.
In \autoref{neutron_skin_plot}, we present the posterior distributions of $\rm r_\mathrm{skin}$ for $^{48}\mathrm{Ca}$ and $^{208}\mathrm{Pb}$ and compare them with the experimental results of the Calcium Radius Experiment (CREX)\cite{CREX:2022kgg} for $^{48}\mathrm{Ca}$ ($0.121\pm 0.026\pm 0.024$ fm) and the Lead Radius Experiment (PREX)\cite{PREX:2021umo} for $^{208}\mathrm{Pb}$ ($0.283 \pm 0.071$ fm). This section discusses the trends in these distributions, the effect of $\chi$-EFT constraints, and their implications for nuclear and astrophysics. 
 
\subsubsection{Neutron Skin Thickness of $^{48}\mathrm{Ca}$}
The top panels of \autoref{neutron_skin_plot} shows the posterior distributions of $\rm r_\mathrm{skin}$ for $^{48}\mathrm{Ca}$. In the fully constrained model ( +NICER 5), the distribution peaks sharply at approximately 0.17 fm, with a narrow 68\% CI. This value is larger than the CREX experimental mean of approximately 0.12 fm, which has a $1\sigma$ range of 0.07--0.17 fm. The median $\rm r_\mathrm{skin}$ of the +NICER 5 distribution coincides with the upper bound of the CREX $1\sigma$ interval, with roughly half of the distribution below the median falling within this range, indicating consistency with the experimental result though with a slight upward shift in the predicted $\rm r_\mathrm{skin}$.
For the case excluding $\chi$-EFT constraints (without $\chi$-EFT), the distribution peaks  approximately at the same value, but with a comparatively broader 68\% CI. 
Also, now more than half of the 68\% CI falls in the $1\sigma$ range of CREX. The slight upward shift relative to the CREX mean in both cases may be attributed to the influence of astrophysical constraints. 

\subsubsection{Neutron Skin Thickness of $^{208}\mathrm{Pb}$}
The posterior distributions for the $\rm r_\mathrm{skin}$ of $^{208}\mathrm{Pb}$ are shown in the bottom panels of \autoref{neutron_skin_plot}. In the fully constrained model (+NICER 5), the distribution peaks sharply at approximately 0.16 fm, with a narrow 68\% CI. This value is notably lower than the PREX experimental mean, which is centered around 0.28 fm with a broad $1\sigma$ range of 0.21--0.35 fm. In the model excluding $\chi$-EFT constraints, the distribution peaks at approximately 0.14 fm, with a comparatively broader 68\% CI. This difference indicates that $\chi$-EFT has significant impact on the predicted $\rm r_\mathrm{skin}$ for $^{208}\mathrm{Pb}$. The 68\% CI lies well outside of the PREX $1\sigma$ range in both the cases, suggesting a potential tension between the model predictions and the PREX result. 
As discussed in the previous section, astrophysical observations favor a softer EoS that correlates with a smaller $\rm r_\mathrm{skin}$ in heavy nuclei like $^{208}\mathrm{Pb}$.

This tension perhaps indicates limitations in the RMF model. This is well documented  in the literature and several attempts have been made to resolve this tension within RMF framework\cite{PhysRevC.109.035803,Miyatsu:2023lki,Kumar:2024bud,Zhao:2024gjz}, by incorporating $\delta$ meson, along with the $\sigma,\omega$ and $\rho$ mesons, as a mediator of the nuclear force. However, none of these has succeeded in satisfying the CREX and PREX skin measurements and all the astrophysical constraints simultaneously. 
\textcite{Zhao:2024gjz} introduced isovector spin-orbit coupling into the RMF description, and by tuning the corresponding parameters they were able to satisfy both the skin measurements and the astrophysical constrains, though only at the cost of an unacceptably large symmetry energy ($E_{\rm sym}\gtrsim 40$ MeV). They also encountered a breakdown of the standard quadratic behavior of symmetry energy with isospin asymmetry leading to different $E_{\rm sym}$ values for almost symmetric nuclear matter and pure neutron matter.
The large uncertainty in the measurement of $\rm r_{\rm skin}^{208}$ make it less effective in constraining the symmetry energy \cite{Zhang:2022bni}. 
However, the future MREX experiment\cite{becker2018p2} or RES-NOVA experiment\cite{pattavina2020neutrino,pattavina2021res,huang2022supernova} can potentially resolve this CREX-PREX tension.

\begin{figure*}[h] 
    \centering
    \subfigure[]{  
        \includegraphics[width=1\columnwidth]{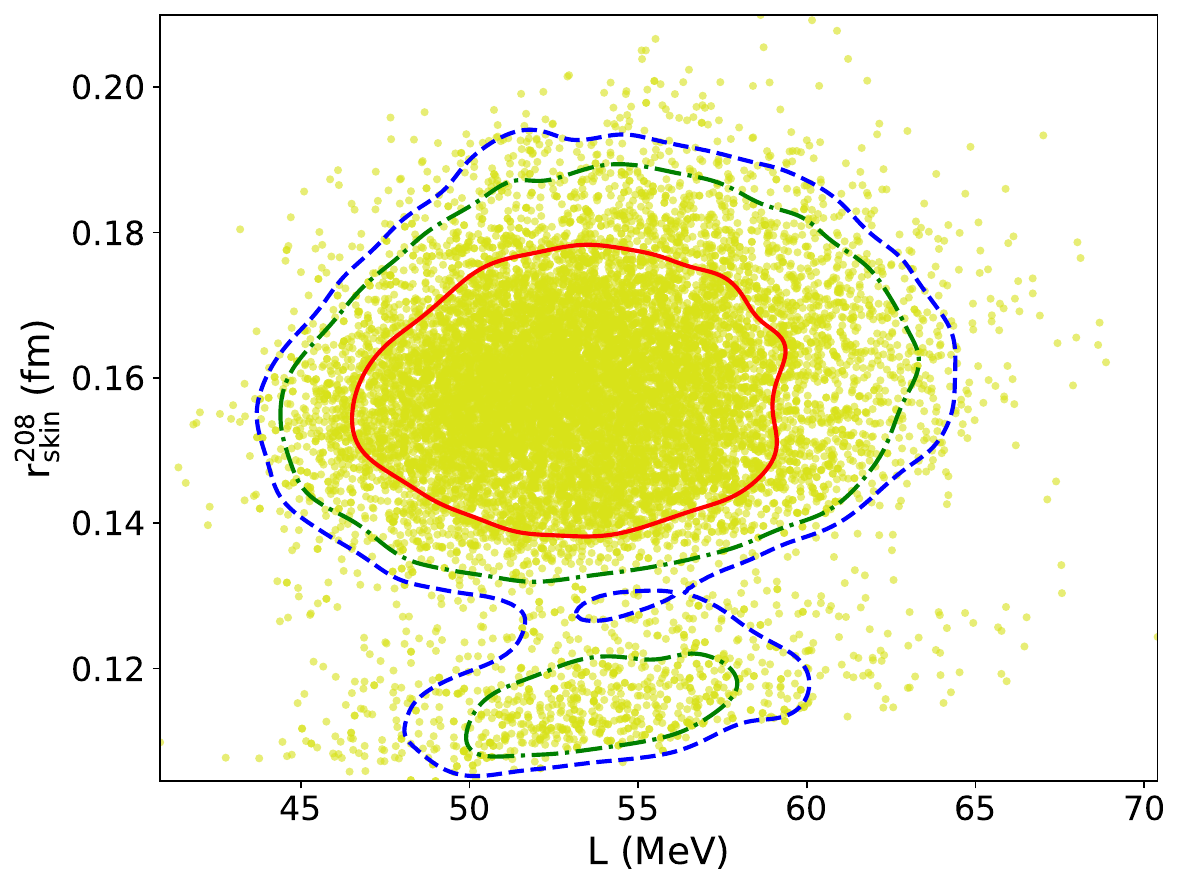}
        \label{fig:skinvsL}
    }
    \hfill
    \subfigure[]{  
        \includegraphics[width=1\columnwidth]{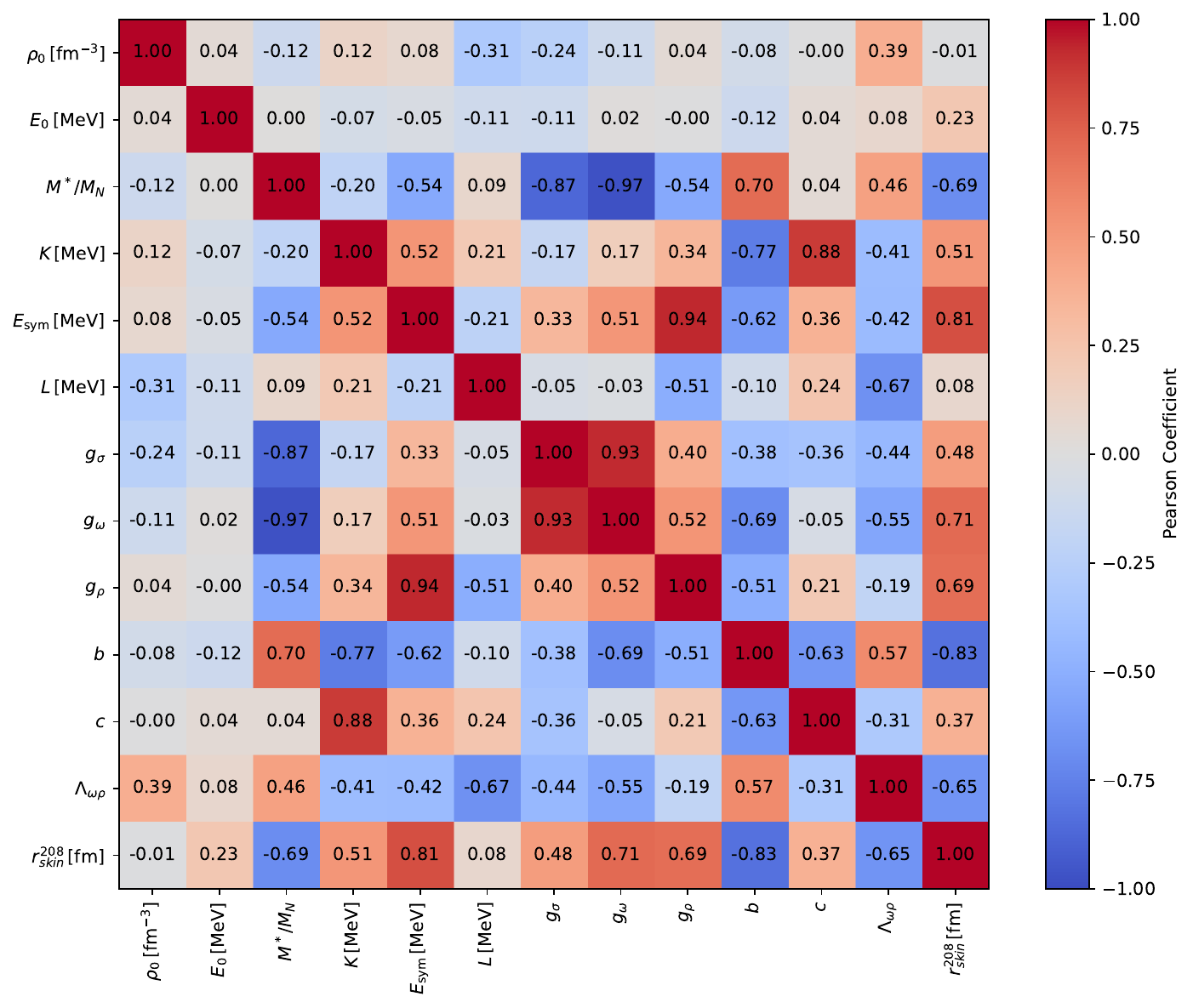}
        \label{fig:corr_Pball+j0614}
    }
    \caption{ (a) Neutron skin thickness of $^{208}$Pb as a function of $L$. The red (solid), green (dash dot), and blue(dashed) contours show the 68\%, 90\%, and 95\% credible regions, respectively. (b) Pearson coefficient matrix for the subset of the posterior that produce a neutron skin for $^{208}\mathrm{Pb}$ nuclei.}
    \label{fig:skinvsL&correlation}
\end{figure*}
Finally, in Figure \autoref{fig:skinvsL}, we present $\rm r_\mathrm{skin}$ of $^{208}$Pb as a function of $L$, using the posterior samples obtained from our Bayesian analysis that incorporates both the $\chi$-EFT inputs and all astrophysical constraints. Interestingly, the results do not exhibit any correlation between these two quantities. This finding contrasts with earlier studies, which reported a strong correlation between  $\rm r_\mathrm{skin}$ and $L$ \cite{Vinas:2013hua,Essick:2021kjb}. This is further supported by Figure \autoref{fig:corr_Pball+j0614}, which shows the Pearson correlation matrix for the saturation parameters and couplings with $\rm r^{208}_\mathrm{skin}$ for the subset of the posterior that produce a neutron skin for $^{208}\mathrm{Pb}$. The correlation coefficient between $L$ and $\rm r_{skin}^{208}$ comes out to be 0.08, effectively indicating no correlation.

The absence of any correlation between $L$ and $\rm r^{208}_\mathrm{skin}$ motivates us to investigate the impact of incorporating neutron skin calculations directly into the likelihood function. However, this turns out to be computationally expensive. Instead, we reweight our posterior samples using the CREX and PREX skin measurements, as is done in literature\cite{PRADHAN2023122578,Ghosh:2021bvw}. This avoids the need to rerun a computationally expensive Bayesian analysis, while still allowing us to assess the impact of including both skin measurements in the likelihood. We show the corner plots for the subset of posterior distributions of our final constraint set when reweighted with CREX and PREX skin measurements(see \autoref{fig:skinweighted_saturation_plus_skins} in Appendix). Even after reweighting, we find no clear correlation. See Appendix for details.

\section{Conclusion}
\label{Conclusion}

In this study, we employed a RMF model to investigate and constrain the EoS of neutron stars. By sequentially applying constraints from chiral effective field theory ($\chi$-EFT), the maximum observed neutron star mass (M$_{\text{max}}$), GW170817, and NICER measurements, we have obtained tight posterior distributions for both nuclear and astrophysical observables, providing a comprehensive picture of the dense nuclear matter.

The sequential application of constraints reveals a clear trend. Within the baseline interval set by $\chi$-EFT and $M_{\rm max}$, GW170817 favors a softer EoS. While NICER 1 and 2 do not significantly shift the median values, they play an important role in constraining the uncertainties of the parameters and observables. NICER 3 and 5 reinforce the preference for a softer EoS, with NICER 5 providing the most stringent constraints.

The RMF model parameters, particularly those governing high-density behavior, are well-constrained. The nuclear saturation parameters exhibit tight posterior distributions, as shown in Table \ref{tab:posterior_ranges_saturation_90CI}. Since the coupling constants are derived from these saturation properties, they are similarly well-constrained (Fig. \ref{fig:corner_couplings}). A significant finding is the growing importance of the $\omega$-$\rho$ coupling ($\Lambda_{\omega \rho}$), which increases from \rn{0.041} to 0.054 as constraints are added (Fig. \ref{fig:corner_couplings}). This parameter, which governs the density dependence of the symmetry energy, becomes increasingly critical under astrophysical constraints, particularly NICER 5, highlighting its role in describing neutron-rich matter in neutron star cores.

The core-crust transition density remains robust under all constraints, while the transition pressure shows a mild non-monotonic variation with steady decrease in uncertainty. 
Also, the transition pressure and density does not show any direct correlation with EoS stiffness, highlighting the dominant role of symmetry energy expansion coefficients and incompressibility. A more detailed investigation of these correlations will be explored in future work.

We also use RMF posterior samples to calculate $\rm r_\mathrm{skin}$ for $^{48}\mathrm{Ca}$ and $^{208}\mathrm{Pb}$, which is a novelty of our work. We then compare our reults with CREX and PREX skin measurements.
For $^{48}$Ca, the model predicts a $\rm r_\mathrm{skin}$ consistent with the CREX measurement in both cases, whether $\chi$-EFT constraints are included or not, reinforcing the model’s robustness in reproducing this experimental result. For $^{208}\mathrm{Pb}$, the predicted skin thickness is significantly lower than the PREX mean of 0.28 fm. When \(\chi\)-EFT is excluded, the peak shifts to 0.14 fm, with the tail of the distribution falling within the PREX 1\(\sigma\) range. This discrepancy in the neutron skin suggests potential tension between nuclear (PREX) and astrophysical (NICER, GW170817) constraints, possibly due to limitations in the RMF model. 
Furthermore, unlike previous studies, we did not find any clear correlation between $\rm r_\mathrm{skin}$ of $^{208}$Pb and the symmetry energy slope $L$, which is one of the highlights of this work.

The pronounced softness of the EoS, driven by GW170817 and NICER 5, may indicate the presence of exotic degrees of freedom such as hyperons and quarks \cite{Nandi:2017rhy, Nandi:2018ami} in neutron star cores, which we plan to explore in future work within our model. The tension between the $^{208}\mathrm{Pb}$ $\rm r_\mathrm{skin}$ and PREX results warrants further investigation. One possibility is to include more terms in the expansion of symmetry energy, which we plan to study next. 

In conclusion, this study reestablishes the power of combining nuclear theory with multi-messenger astrophysics to constrain the neutron star EoS and nuclear properties. The tight constraints on model parameters, the softening of the EoS, and the successful reproduction of CREX highlight the robustness of our approach, while the discrepancy with PREX opens new avenues for research. Future gravitational wave and NICER observations, upcoming laboratory measurements such as MREX, and supernova neutrino detection efforts like RES-NOVA could further tighten these constraints, enhancing our understanding of dense matter physics.

\section*{Acknowledgements}
The authors acknowledge the High Performance Computing Cluster (HPCC) 'Magus' at Shiv Nadar Institution of Eminence and Tejas High Performance Computing Cluster (HPCC) at IIT Delhi for providing computational resources that have contributed to the research results reported within this paper. RK gratefully acknowledges the Council of Scientific and Industrial Research (CSIR), India, for providing financial support in the form of a Junior Research Fellowship (JRF) under the award number 09/0086(24152)/2025-EMR-I. This project has received funding from the European Union’s Horizon 2020 research and innovation programme under the Marie Skłodowska-Curie grant agreement No. 101034371. PC acknowledges the support from the European Union's HORIZON MSCA-2022-PF-01-01 Programme under Grant Agreement No. 101109652, project ProMatEx-NS. P.C. acknowledges further support from the grant PID2024-158938NB-I00 funded by MICIU/AEI/10.13039/501100011033 and by "ERDF A way of making Europe."

\bibliography{mybiblio}

\appendix
\section*{Appendix}
\label{Appendix}

\begin{figure*}
    \centering
    \includegraphics[width=1\linewidth]{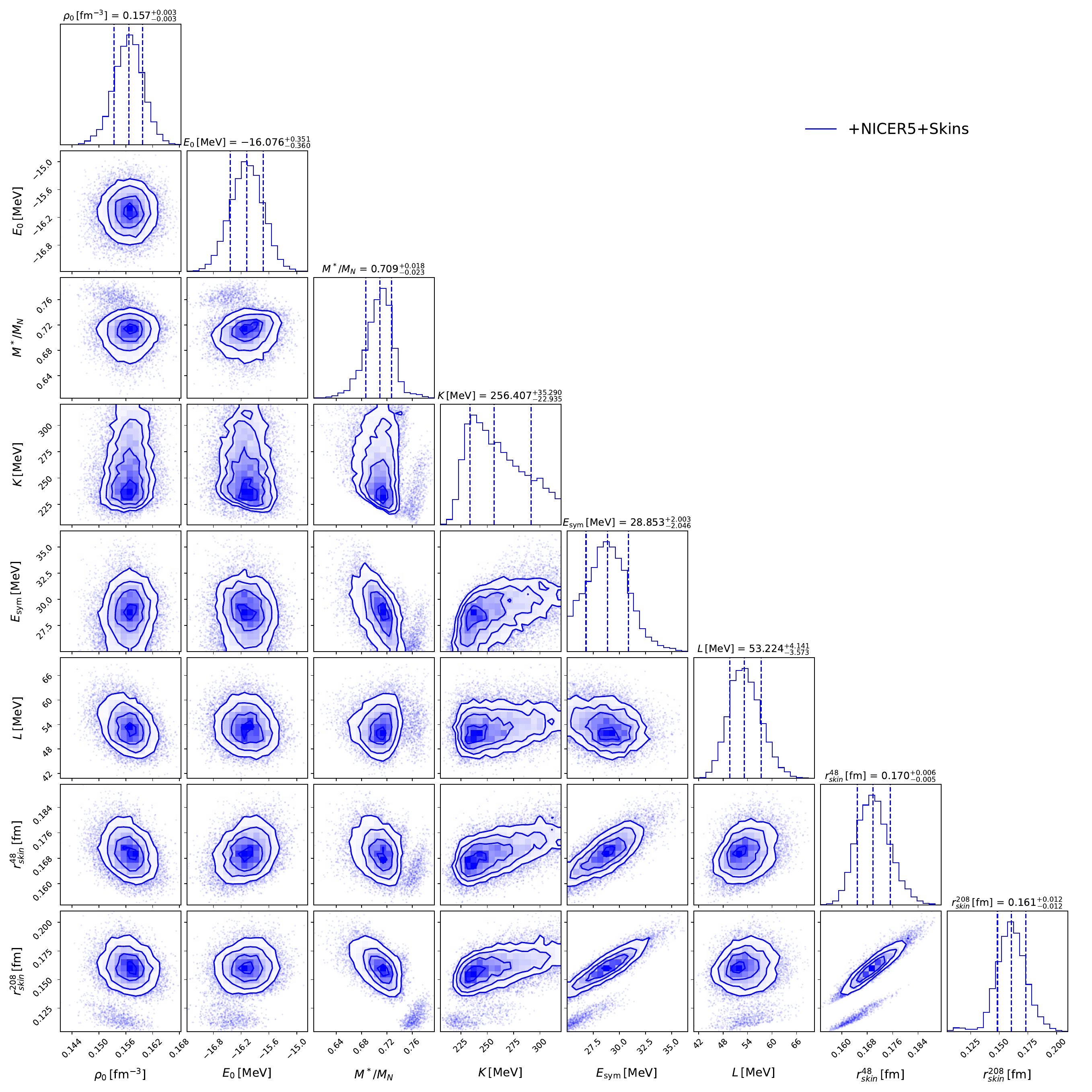}
    \caption{Corner plot of the posterior distributions obtained by reweighting our final constraint set (+NICER 5) with the neutron skin thickness measurements of $^{48}\mathrm{Ca}$ (CREX) and $^{208}\mathrm{Pb}$ (PREX), showing the nuclear saturation parameters and skin thicknesses of both nuclei. This reweighted posterior is referred to as +NICER5+Skins. Diagonal panels display the one-dimensional marginalized posteriors for each parameter, with the 68\% CI shown above. Off-diagonal panels show the two-dimensional joint probability density functions between parameter pairs.}
    \label{fig:skinweighted_saturation_plus_skins}
\end{figure*}

\begin{figure*}
    \centering
    \includegraphics[width=1\linewidth]{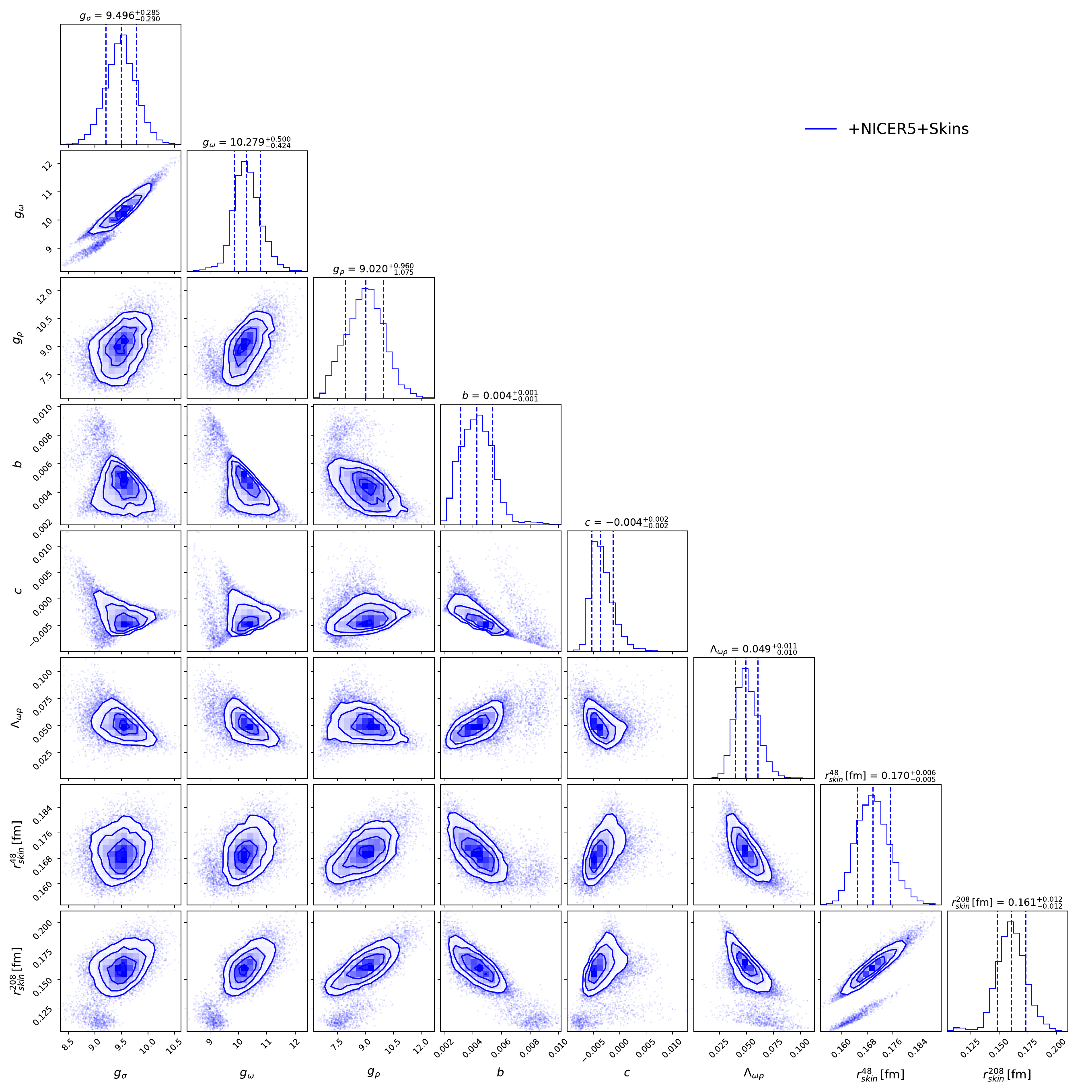}
    \caption{Same as \autoref{fig:skinweighted_saturation_plus_skins} but now for coupling constants and skins of both nuclei.}    
    \label{fig:skinweighted_couplings_plus_skins}
\end{figure*}

\begin{figure*}  
    \centering
    \subfigure[]{  
        \includegraphics[width=1\columnwidth]{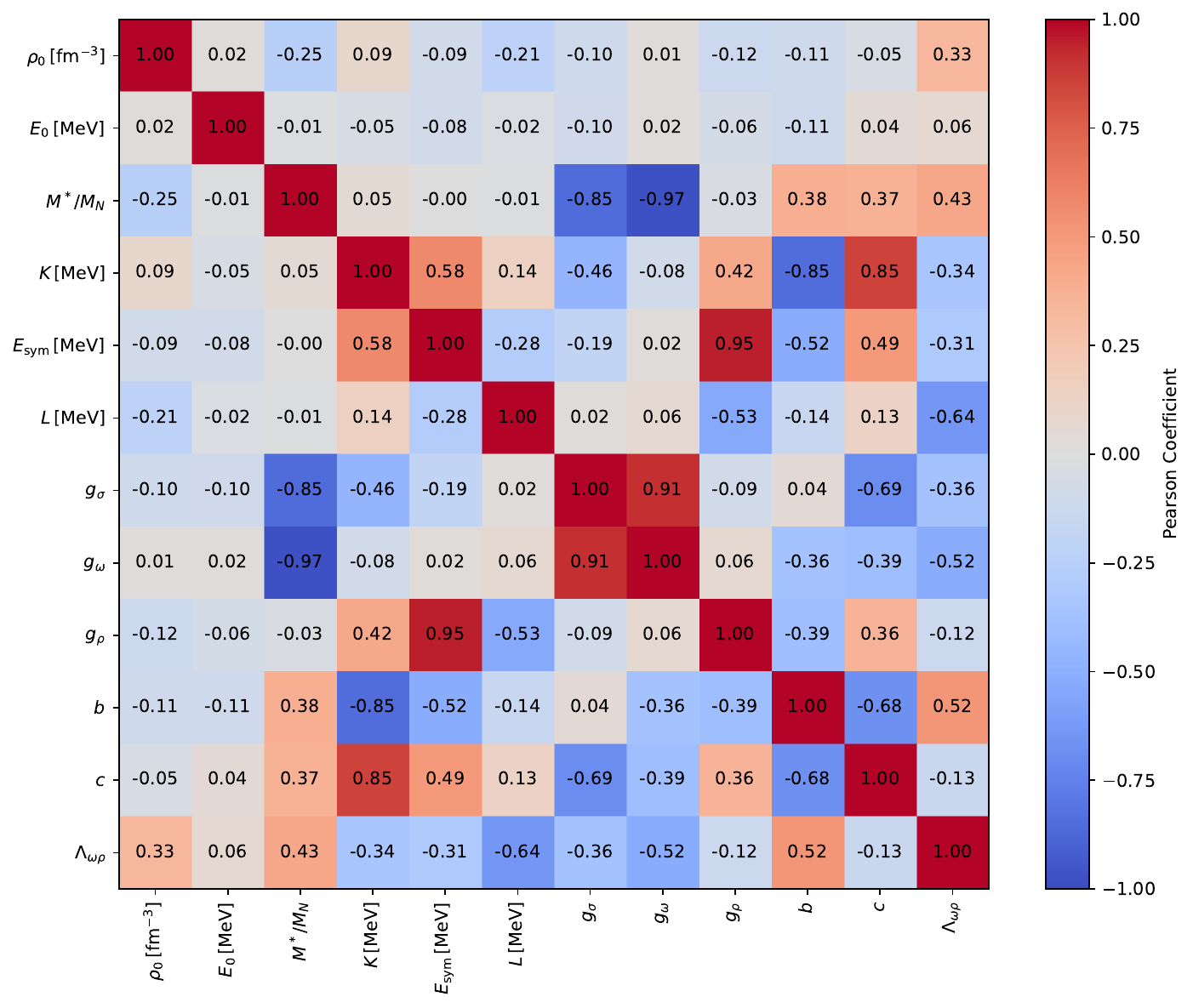}
        \label{fig:corr_matrix_final}
    }
    \hfill
    \subfigure[]{  
        \includegraphics[width=1\columnwidth]{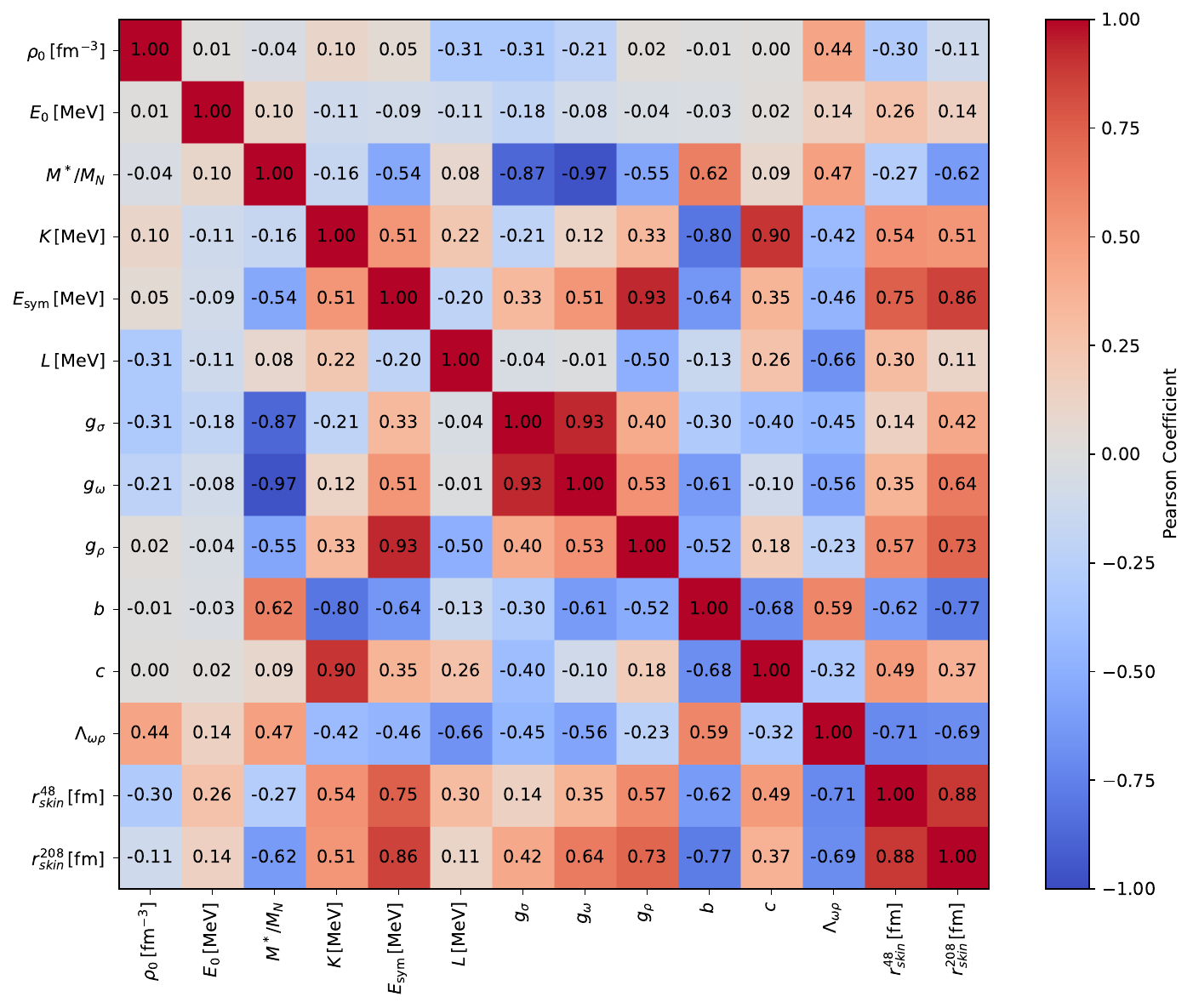}
        \label{fig:corr_matrix_skin_weighted}
    }
    \caption{Pearson correlation matrices before (left, +NICER 5) and after (right, +NICER5+Skins) reweighting with the neutron skin thickness measurements of $^{48}\mathrm{Ca}$ (CREX) and $^{208}\mathrm{Pb}$ (PREX). The color scale indicates the strength and sign of the correlation.}
    \label{fig:correlation_heatmap}
\end{figure*}

The inclusion of CREX and PREX constraints in the likelihood calculation is computationally very expensive as that involves calculation of $\rm r_\mathrm{skin}$ of calcium and lead for each parameter set. Instead, we reweight our equally-weighted posterior samples for our final constraint set(+NICER 5) with CREX and PREX 1$\sigma$ interval. This is computationally very efficient and also helps us to see the effect of skin constraints on our parameters and couplings. 
For achieving this, first we calculate $\rm r_\mathrm{skin}$ for both $^{48}\mathrm{Ca}$ and $^{208}\mathrm{Pb}$ nuclei. During the process, not every parameter set is able to converge and provide us with a neutron skin. The samples which do not provide us with a finite neutron skin for either of the nuclei are discarded.  
Then we construct Gaussians from 1$\sigma$ interval of CREX and PREX and assign weights to the remaining parameter sets. After that we plot the weighted posterior (denoted as +NICER5+Skins) into a corner plot (see \autoref{fig:skinweighted_saturation_plus_skins} and \autoref{fig:skinweighted_couplings_plus_skins}). We then calculate the Pearson correlation coefficient for the final constraint set(+NICER 5) and the reweighted posterior(+NICER5+Skins). \autoref{fig:correlation_heatmap} shows the correlation heatmaps corresponding to +NICER 5 (left) and +NICER5+Skins (right) with CREX and PREX skin measurements. 
Since not all parameter sets produce $\rm r_{skin}$ for both nuclei, correlations can't be calculated with $\rm r_{skin}^{48}$ and $\rm r_{skin}^{208}$. This is the reason these quantities are absent in the left panel.

\end{document}